\documentclass[12pt]{article}
\usepackage[a4paper, margin=1in]{geometry} 
\usepackage{times}       

\usepackage{setspace}
\usepackage{appendix}
\usepackage{amsmath}
\usepackage{graphicx}
\usepackage{natbib}
\usepackage{url} 
\usepackage{amssymb}
\usepackage{amsthm}
\usepackage{babel}[USenglish, american]
\usepackage{comment}
\usepackage{xcolor}

\usepackage{tabularx}
\usepackage{algorithm}
\usepackage{algpseudocode}

\newcommand{\E}{\ensuremath{\mathrm{E}}}

\newcommand{\bk}{\mathbf{k}}

\newcommand{\bs}{\mathbf{s}}

\newcommand{\bh}{\mathbf{h}}

\newcommand{\bv}{\mathbf{v}}

\newcommand{\cov}{\ensuremath{\mathrm{cov}}}

\newtheorem{proposition}{Proposition}[section]
\newtheorem{definition}{Definition}[section]

\usepackage{comment}
\usepackage{graphicx}

\usepackage{authblk}

\usepackage{tikz}
\usetikzlibrary{shapes.geometric, arrows}

\tikzstyle{hom_model} = [rectangle, rounded corners, 
minimum width=3cm, 
minimum height=1cm,
text centered, 
draw=black, 
fill=blue!30]

\tikzstyle{het_model} = [rectangle, rounded corners, 
minimum width=3cm, 
minimum height=1cm,
text centered, 
draw=black, 
fill=red!30]

\tikzstyle{hp} = [rectangle, 
minimum width=3cm, 
minimum height=1cm, 
text centered, 
draw=black, 
fill=green!30]
\tikzstyle{arrow} = [thick,->,>=stealth]

\begin{document}
\emergencystretch 3em

\author[1]{Maria Laura Battagliola}
\author[2]{Sofia C. Olhede}
\affil[1]{Department of Statistics, ITAM, Río Hondo 1, Altavista, Álvaro Obregón, 01080 Mexico City, Mexico}

\affil[2]{Institute of Mathematics, EPFL, Station 17, 1015 Lausanne, Vaud, Switzerland}
{
    \makeatletter
    \renewcommand\AB@affilsepx{: \protect\Affilfont}
    \makeatother

    \affil[ ]{Contacts}

    \makeatletter
    \renewcommand\AB@affilsepx{, \protect\Affilfont}
    \makeatother

    \affil[1]{laura.battagliola@itam.mx}
    \affil[2]{sofia.olhede@epfl.ch}
}

\title{\bf Modeling Spatio-Temporal Transport: From Rigid Advection to Realistic Dynamics}

  \maketitle

\bigskip
\begin{abstract}

Stochastic models for spatio-temporal transport face a critical trade-off between physical realism and interpretability. The advection model with a single constant velocity is interpretable but physically limited by its perfect correlation over time. This work aims to bridge the gap between this simple framework and its physically realistic extensions. Our guiding principle is to introduce a spatial correlation structure that vanishes over time. To achieve this, we present two distinct approaches. The first constructs complex velocity structures, either through superpositions of advection components or by allowing the velocity to vary locally. The second is a spectral technique that replaces the singular spectrum of rigid advection with a more flexible form, introducing temporal decorrelation controlled by parameters. We accompany these models with efficient simulation algorithms and demonstrate their success in replicating complex dynamics, such as tropical cyclones and the solutions of partial differential equations. Finally, we illustrate the practical utility of the proposed framework by comparing its simulations to real-world precipitation data from Hurricane Florence.

\end{abstract}

\noindent%
\vfill
\vfill

\newpage

\section{Introduction}
\label{sec:intro}

Spatio-temporal phenomena driven by advection are fundamental to a vast array of scientific disciplines, from tracking atmospheric pollutants and forecasting ocean currents to modeling seismic wave propagation. The modeling of these processes is challenging due to their complex interdependent evolution in space and time.
While the field of spatio-temporal statistics has made substantial progress, with seminal works like \cite{Cressie2011}, accurately capturing the dynamics of transport remains an active and critical area of research.

To model these dynamics, one major research direction uses Lagrangian approaches to build non-stationary covariance structures by tracking particle trajectories. This framework has proven effective for modeling environmental processes like ocean waves and precipitation \citep{ailliot2011, benoit2018}, as these phenomena are inherently governed by physical transport.
Another line of research employs Stochastic Partial Differential Equations (SPDEs) to embed physical principles like advection and diffusion directly into the models. This has led to computationally efficient models with interpretable parameters \citep{sigrist2015}, which have since been extended to handle more complex, spatially-varying dynamics \citep{liu2022, carlotto2024}. While other methods focus on simulating broader classes of non-separable covariance functions (see for instance \cite{allard2020simulating}), they often face a trade-off: they either omit explicit directional transport for simplicity, or incorporate it at the cost of significant model complexity. Our work seeks to occupy a middle ground, prioritizing both physical realism and interpretability.
For comprehensive reviews on the state of dynamic and non-separable models, namely models that allow spatial and temporal dependence 
to interact, we refer to
\cite{gneiting2002nonseparable}, \cite{porcu2020nonseparable}, \cite{chen2021space}, and \cite{papalexiou2021advancing}. 

To achieve physical realism, our work begins with the foundational model of a spatial random field advected by a constant velocity—a classic benchmark used to approximate natural phenomena in works like \cite{cox1988}. Our goal is to develop extensions from this starting point that add complexity while retaining the original's intuitive interpretability.
We refer to the simplest motivating model as the {\em frozen field}, following the terminology of \cite{christakos2017}. The advantage of this starting point is its structure. In particular, its value, covariance, and spectrum are all defined by three simple components: the initial spatial field, its covariance function, and the constant velocity. This allows us to systematically manipulate these components to build more complex and realistic models while maintaining the clear physical interpretation of the original. 

To this end, we explore two distinct strategies to generalize the frozen field. The first strategy focuses on enriching the velocity structure itself, moving beyond a single constant vector. 
This leads to two models, the first of which, related to earlier work on sums of advected fields \citep{schlather2010some}, relies on the superposition of multiple frozen fields, while the second considers a velocity that is itself a vector field varying in space and time.
Our second strategy involves relaxing the spectral shape of the frozen field, replacing its singular frequency-domain representation with more flexible forms that allow for natural decorrelation over time, which can be controlled with parameters in the model.

To complement our theoretical analysis, we also develop efficient simulation methods, which are vital for model validation, parameter inference, and prediction. Our approach to simulation mirrors our theoretical strategy: we first establish methods for the frozen field model, and then extend them to more complex frameworks. Specifically, we begin by introducing two distinct procedures to simulate a frozen field, one in the space-time domain and the other in the spectral domain. We then demonstrate how these algorithms can be adapted to simulate the novel models proposed in this work.

Thus, this paper makes a threefold contribution. First, we introduce these novel classes of models for advection. Second, we provide computationally efficient algorithms to simulate them, ensuring that these methods are accessible for practical application. We demonstrate that these algorithms are competitive with, and in some cases superior to, standard approaches by avoiding known issues like numerical diffusion, namely the artificial spreading caused by discretization schemes. Third, we showcase the applicability of these models by illustrating their ability to generate complex, physically plausible dynamics, such as those of a tropical cyclone, and to replicate the solutions of well-known transport and diffusion PDEs. We further validate our approach with a real-data application to Hurricane Florence, comparing our simulations directly to observational precipitation records.
To facilitate their use by practitioners, we provide implementations of these methods in \texttt{R} \citep{Rmanual}.

The paper is organized as follows. Section 2 formally defines the frozen field model and discusses two primary simulation approaches. 
Section 3 introduces our first class of generalizations based on velocity structures, presents simulated examples, compares the results to numerical PDE solvers, and assesses the predictive performance against observed data from Hurricane Florence.
Section 4 details our second generalization strategy based on spectral modifications and analyzes the role of the model parameters. Finally, Section 5 provides concluding remarks and outlines directions for future research. Further results are included in the supplementary materials.

\section{Space Random Fields with Advection}
\label{sec:background}

In what follows, we give a formal definition of a spatial random field with advection, including the analytical formula of its covariance and spectral density functions. Then, we propose two numerical algorithms to efficiently simulate this class of phenomena.

\subsection{Framework and definition}
\label{sec:background_theory}

We begin by defining fundamental concepts for spatial random fields, which form the basis for the spatio-temporal models discussed later. Consider a continuous-space random field $X_S: \mathbb{R}^d \to \mathbb{R}$, where $X_S(\bs)$ is the value of the field at the spatial coordinate $\bs \in \mathbb{R}^d$. For simplicity, we consider $d=2$, though extensions to higher dimensions are generally straightforward. Assuming $X_S(\bs)$ is a Gaussian random field, its probabilistic structure is fully characterized by its mean function $\E\{X_S(\bs)\}$ and its covariance function $c_{XX}(\bs_1, \bs_2) = \cov\{X_S(\bs_1), X_S(\bs_2)\}$ for any spatial locations $\bs,\bs_1, \bs_2 \in \mathbb{R}^2$.

A common and important simplifying assumption is spatial stationarity. A spatial random field $X_S(\bs)$ is said to be stationary if its mean is constant (assumed to be zero without loss of generality for covariance modeling) and its covariance function $c_{XX}(\bs_1, \bs_2)$ depends only on the lag $\mathbf{h} = ( \bs_2 - \bs_1)\in \mathbb{R}^2$. With a slight abuse of notation, we write this stationary covariance function as
\begin{align*}
    c_{XX}(\mathbf{h}) = \cov\{X_S(\bs), X_S(\bs + \mathbf{h})\}.
\end{align*}
 Under stationarity, the spectral analysis of $X_S(\bs)$ becomes particularly useful. The $2$-dimensional Fourier transform of the stationary covariance function $c_{XX}(\mathbf{h})$ yields the spectral density function (or spectrum), $S_{XX}(\mathbf{k})$:
\begin{align*}
    S_{XX}(\mathbf{k}) = \int_{\mathbb{R}^2} c_{XX}(\mathbf{h}) e^{-2\pi i \mathbf{k}^\top \mathbf{h}} \mathrm{d}\mathbf{h},
\end{align*}
where $\mathbf{k} \in \mathbb{R}^2$ is the wavenumber. Being the Fourier transform of the covariance function, the spectrum $S_{XX}(\mathbf{k})$ describes the contribution of different frequencies to the field's variance. \\

Having established the properties of spatial random fields, we now extend these concepts to the space-time domain to model evolving phenomena. A space-time random field $Z: \mathbb{R}^2 \times \mathbb{R}^+ \to \mathbb{R}$ takes values over space-time coordinates $(\bs, t) \in \mathbb{R}^2 \times \mathbb{R}^+$.
We consider a special class of space-time random fields, namely that of \textit{traveling random fields}. One can think of them as space random fields advected over time. The simplest form for such phenomena 
consists in a spatial field travelling with constant velocity $\bv \in \mathbb{R}^2$. This entails that the random field is ``frozen" and rigidly shifted in time according to the velocity. We adopt the name \textit{frozen field} from \cite{christakos2017spatiotemporal} to denote such space-time process.
We recall the definition of this stochastic process.
\begin{definition}[Frozen Field]
\label{def:frozen}
A random field $Z(\bs,t)$ is a {\em Frozen Field} if there is a 
constant velocity $\mathbf{v}\in{\mathbb{R}}^2$, and a random field $X_S(\bs)$ with $\bs\in {\mathbb{R}}^2$
and covariance function $c_{XX}(\bh)$ such that
\begin{align}
Z(\bs,t)=X_S(\bs-\mathbf{v} t),
\label{eq:frozen_field}
\end{align}
with $t \in \mathbb{R}^+$. 
\end{definition}
The Frozen Field (FF from here onward) $Z(\bs,t)$ is the solution of transport equation
\begin{align}
\label{eq:transport_PDE}
\begin{cases}
     \frac{\partial Z(\bs,t)}{\partial t} + \bv \cdot \nabla Z(\bs,t) = 0,\\
     Z(\bs,0) = X_S(\bs),
\end{cases}
\end{align}
where $\nabla Z(\bs,t) = \left( \frac{\partial  Z(\bs,t)}{\partial s_1} \frac{\partial  Z(\bs,t)}{\partial s_2} \right)^\top$. Moreover, this process has the following second order properties:
\begin{align}
\label{eq:ff_cov}
c_{ZZ}(\bh,\tau)&=    c_{XX}(\bh-\mathbf{v} \tau),\\
S_{ZZ}(\bk,\omega)&=S_{XX}(\bk)\cdot \delta (\omega+\bk^\top \mathbf{v}),
\label{eqn:ff_spectrum}
\end{align}
where $(\bk, \omega) \in \mathbb{R}^2 \times \mathbb{R}$ denotes the pair of wavenumber and frequency, respectively.
Specifically, the covariance $c_{ZZ}(\bh,\tau)$ consists of the covariance function $ c_{XX}(\cdot)$ of the space random field evaluated at spatial lag $(\bh-\mathbf{v} \tau)$. This structure 
makes the FF model non-separable, as the spatial and temporal dependencies 
are intrinsically coupled through the advection term $\mathbf{v}\tau$.
The spectrum $S_{ZZ}(\bk,\omega)$, i.e. the Fourier transform of the covariance function $c_{ZZ}(\bh,\tau)$,
 depends on the spectrum of the space random field $S_{XX}(\bk)$, which is multiplied by a generalized function corresponding to a Dirac delta depending on velocity $\mathbf{v}$. 

In this space-time model, spatial locations shift according to the constant velocity as time progresses. More specifically, spatial correlation $\rho(\bh, \tau) = \frac{\left | c_{XX}(\bh-\mathbf{v} \tau) \right |}{c_{XX}(0)}$ of the FF is such that
\begin{align*}
    \rho(\bv \tau, \tau) = \frac{ c_{XX}(\bv \tau-\mathbf{v} \tau)}{c_{XX}(0)} = 1,
\end{align*}
namely the correlation is perfect along {\em propagation path} $(\bv \tau, \tau)$. However, under normal circumstances, one would not expect an advection phenomenon to persist over all positive lags, as this would imply a particle traveling through a perfectly stationary medium without friction. 
Energy dissipation, namely the loss of kinetic energy over time, would be a more realistic expectation \citep{christakos2017spatiotemporal}.
On the other hand, despite being overly-simplistic, the FF model enjoys the striking benefit of being a {\em straightforwardly interpretable} model for nonseparable space-time random fields. Relying on such interpretability and driven by the necessity for flexibility in describing complex spatiotemporal phenomena, the core of this work is to introduce new models by relaxing certain assumptions of the FF model. 

As we will provide simulation procedure for the these new models, we first describe two algorithms to simulate a FF. This is the content of the next section.

\subsection{Simulation of Frozen Fields}

Say we wish to simulate a FF traveling with known velocity $\bv$. Such space-time process is discretized over a spatial grid $\mathcal{G} = [0,n] \times [0,n] \in \mathbb{N}_0^2$, which is square for sake of simplicity, and time grid $\{t_0, \dots, t_T\} \in \mathbb{N}_0$. As mentioned previously, this can be done in two ways.

Firstly, a FF can be generated by simulating a single, large spatial random field, and then advecting $\mathcal{G}$ across this field over time. More precisely, we first need to generate the extended spatial field $X_S^{\text{ext}}(\bs)$ on grid $\mathcal{G}^* = [0,N] \times [0,N]$, with $N>n$. This initial simulation can be performed with circulant embedding \citep{ChanWood97, DietrichNewsam97, Graham2018}, a non-exact spectral method relying on the use of Discrete Fourier Transforms (DFTs) and their implementation on regular grids, called Fast Fourier Transforms (FFTs). The computational cost of circulant embedding, for which a FFT is carried out once, is $\mathcal{O}(N^2 \log N)$.
Notice that the simulation domain $\mathcal{G}^*$ must be large enough to contain the full trajectory of the smaller observation grid $\mathcal{G}$. The required grid size $N$ should be chosen based on the properties of the spatio-temporal field being simulated. Accordingly, we set $N = n + T v_{\text{max}}$, where $v_{\text{max}} = \max\{|v_1|, |v_2|\}$ is the maximum component of the constant velocity $\bv = (v_1, v_2)^\top $.

Once $X_S^{\text{ext}}$ is generated, the space-time field $Z(\bs, t)$ is constructed by sampling from this single large field. For each point $\bs$ on the observation grid $\mathcal{G}$, and for each of the $T$ time steps, the value is given by $Z(\bs, t) = X_S^{\text{ext}}(\bs - \bv t)$. It is important to note that the advected location $\bs - \bv t$ will generally not coincide with a point on the discrete grid $\mathcal{G}^*$. Therefore, its value must be obtained through a spatial interpolation scheme, such as bilinear or bicubic interpolation. This sampling and interpolation process for all points and time steps has a total cost of $\mathcal{O}(n^2 T)$.

A comparison of computational costs reveals a clear trade-off between this shifting method and a direct space-time simulation via 3D circulant embedding, which costs $\mathcal{O}(n^2 T \log(nT))$. The total cost of the shifting method, $\mathcal{O}(N^2 \log N + n^2 T)$, depends heavily on $N$, which, by construction, grows linearly with $T$. Consequently, the shifting method is significantly more efficient when the number of time steps $T$ is small relative to the spatial grid dimension $n$. For long $T$, the 3D spectral approach becomes more advantageous as its cost scales less aggressively with time. For context, both methods are vastly more efficient than an exact simulation using Cholesky decomposition of the full space-time covariance matrix, which would require a cost of $\mathcal{O}(n^6 T^3)$. The overall procedure for the shifting method is detailed in Algorithm \ref{alg:sim_FF}.

\begin{algorithm}[H]
\caption{Simulation of a Frozen Field via Grid Shifting}\label{alg:sim_FF}
\begin{algorithmic}[1]
\Require Spatial grid dimension $n$, number of time steps $T$, constant velocity $\bv$, spatial covariance function $c_{XX}$.
\State Compute max velocity component $v_{\text{max}} \gets \max\{|v_1|, |v_2|\}$
\State Compute extended grid dimension $N \gets n + T v_{\text{max}}$
\State Define extended grid $\mathcal{G}^{*} \gets [0,N] \times [0,N]$ \Comment{Grid that accommodates for advection}
\State Generate Gaussian random field $X_{S}^{\text{ext}}$ on $\mathcal{G}^{*}$ using $c_{XX}$ \Comment{Simulate large background field}
\State Define simulation grid $\mathcal{G} \gets [0,n] \times [0,n]$
\State Initialize space-time field $Z \in \mathbb{R}^{(n+1) \times (n+1) \times (T+1)}$
\For{each spatial location $\bs$ in $\mathcal{G}$}
    \State $Z(\bs, 0) \gets X_{S}^{\text{ext}}(\bs)$ \Comment{Set initial state from the large field}
    \For{$t = 1$ to $T$}
        \State $Z(\bs, t) \gets X_{S}^{\text{ext}}(\bs - \bv t)$ \Comment{Advect by sampling the large field via interpolation}
    \EndFor
\EndFor
\State \Return $Z$
\end{algorithmic}
\end{algorithm}

A second approach is an adaptation of the spectral method called Davies-Harte algorithm (see for instance \cite{craigmile2003}). This method simulates the Gaussian random field based on its spectral density function. The procedure, detailed in Algorithm \ref{alg:DH}, begins by defining padded grid dimensions, $n_{\text{pad}} = 2n$ and $T_{\text{pad}} = 2T$. This step is crucial for mitigating the periodic boundary effects that are inherent in any FFT-based simulation. The wavenumber and frequency vectors, $k_{vec}$ and $\omega_{vec}$, are defined as regular sequences spanning the fundamental frequency domain, with uniform spacing $1/n_{pad}$ and $1/T_{pad}$ respectively, ensuring compatibility with the discrete Fourier transform.

On this padded grid, a complex-valued field, $\widetilde{Z}$, is constructed. For each point in the discrete frequency domain, the value of $\widetilde{Z}$ is set by multiplying the square root of the space-time spectrum, $\sqrt{S_{ZZ}(\bk, \omega)}$, by a random complex white noise sample. Once this complex grid is fully constructed, it is transformed to the space-time domain using a single 3D Inverse FFT (IFFT). The real part of the resulting padded field is taken, and the final $n \times n \times T$ field is obtained by cropping the result, discarding the extra grid points that were added for padding.

A key practical consideration for this spectral method is its accuracy. Since the theoretical spectrum \eqref{eqn:ff_spectrum} contains a Dirac delta function, which cannot be represented exactly on a discrete frequency grid, it must be approximated, for example, with a narrow Gaussian kernel. This approximation introduces a trade-off. For a given wavenumber-frequency grid resolution, a narrower kernel more accurately represents the Dirac delta, but can lead to higher numerical instability or aliasing. Conversely, a wider kernel smooths the spectrum, deviating more from the advected behavior, but offering greater numerical stability. Furthermore, as mentioned before, to mitigate artificial correlations arising from periodic boundary effects, the Davies-Harte algorithm employs padding by doubling the grid size in all dimensions. This effectively creates a buffer region where the artificial periodicities occur, ensuring that the central, desired portion of the simulated field remains largely free from such distortions. However, the overall quality of the simulation is also influenced by the discrete resolution of the frequency grid, which approximates the continuous theoretical spectrum. The computational cost of this entire procedure is dominated by the single 3D IFFT, leading to a complexity of $\mathcal{O}(n^2 T \log(nT))$.

\begin{algorithm}[H]
\caption{Simulation of a Frozen Field via its Spectrum (Davies-Harte)}\label{alg:DH}
\begin{algorithmic}[1]
\Require Spatial grid dimension $n$, number of time steps $T$, space-time spectral density function $S_{ZZ}$.
\State Define padded dimensions $n_{\text{pad}} \gets 2n$ and $T_{\text{pad}} \gets 2T$ \Comment{Double grid sizes}
\State Define wavenumber vector $\mathbf{k}_{\text{vec}}$ for a grid of size $n_{\text{pad}}$
\Comment Fourier frequencies
\State Define frequency vector $\omega_{\text{vec}}$ for a grid of size $T_{\text{pad}}$
\Comment Fourier frequencies
\State Initialize complex grid in frequency domain $\widetilde{Z} \in \mathbb{C}^{n_{\text{pad}} \times n_{\text{pad}} \times T_{\text{pad}}}$

\For{each index $(i, j, t)$ in the padded grid dimensions}
    \State Get corresponding wavevector $\bk \gets (\mathbf{k}_{\text{vec}}[i], \mathbf{k}_{\text{vec}}[j])$ and frequency $\omega \gets \omega_{\text{vec}}[t]$
    \State Evaluate spectral density $S \gets S_{ZZ}(\bk, \omega)$
    \State Generate complex white noise $\gamma \sim \mathcal{CN}(0,1)$ \Comment{$\Re(\gamma), \Im(\gamma) \sim \mathcal{N}(0,1)$}
    \State $\widetilde{Z}[i,j,t] \gets \sqrt{S} \cdot \gamma$ \Comment{Construct the random field in frequency domain}
\EndFor

\State Perform inverse Fourier transform $Z_{\text{pad}} \gets \text{IFFT3}(\widetilde{Z})$ \Comment{Transform to space-time domain}
\State Take the real part $Z_{\text{pad}} \gets \Re(Z_{\text{pad}})$
\State Crop to original dimensions $Z \gets Z_{\text{pad}}[0:n-1, 0:n-1, 0:T-1]$

\State \Return $Z$
\end{algorithmic}
\end{algorithm}

Having established the properties and simulation methods for the FF model, we now turn to the analysis of its extensions.

\section{Generalization via Velocity Structures}

This section is dedicated to the extensions of the FF model with our first strategy, namely modifying the overly simplistic velocity structure of model \eqref{eq:frozen_field}. In particular, we consider two models, one consisting of the overlapping of FFs, and one where we substitute the constant velocity with a space-time dependent vector field. We study the properties of both of them, including their propagation path. 

\subsection{Framework}
\label{sec:Distr_Ev_FF_framework}

Our first extension of model \eqref{eq:frozen_field} consists of juxtaposition of several fields with different advection directions. To do so, consider geometric transformations of a constant velocity $\mathbf{v} \in \mathbb{R}^{2}$ in \eqref{eq:frozen_field}. Specifically, we examine composite transformations $\mathbf{M} \in \mathbb{R}^{2\times 2}$ of the form
\begin{equation}
    \mathbf{M} = \mathbf{R}\mathbf{S},
    \label{eq:matrix_transformation}
\end{equation}
where $\mathbf{R} \in \mathbb{R}^{2\times 2}$ is a rigid rotation around the origin and $\mathbf{S} \in \mathbb{R}^{2\times 2}$ denotes a scaling transformation.

\begin{definition}[Distributed FF]
\label{def:distrFF}
Consider $n_{\text{v}}$ rotations $\mathbf{R}_1, \dots, \mathbf{R}_{n_{\text{v}}}$ and scalings $\mathbf{S}_1, \dots, \mathbf{S}_{n_{\text{v}}}$. A random field $Z(\bs,t)$ is a {\em Distributed Frozen Field} if, given a velocity $\bv\in \mathbb{R}^2$, there exists a set of transformations $\mathbf{M}_1, \dots, \mathbf{M}_{n_{\text{v}}}$ defined as in \eqref{eq:matrix_transformation}, and a random field $X_S(\bs)$ with $\bs\in \mathbb{R}^2$ and covariance function $c_{XX}(\bh)$ such that

\begin{equation}
\label{eq:DistributedFF_combination}
Z(\bs,t)=\sum_{i=1}^{n_{\text{v}}} p(\mathbf{M}_i\bv) Z_i(\bs,t)=\sum_{i=1}^{n_{\text{v}}} p(\mathbf{M}_i\bv) X_S \left(\bs - (\mathbf{M}_i\mathbf{v})t\right),
\end{equation}

for $t \in \mathbb{R}^+$, where the weights satisfy
\begin{align*}
    \sum_{i=1}^{n_{\text{v}}} p(\mathbf{M}_i \mathbf{v}) &= 1, \\
    p(\mathbf{M}_i \mathbf{v}) &\geq 0, \quad \forall i = 1,\dots,n_{\text{v}}. 
\end{align*}
\end{definition}
The constraints requiring the weights to be non-negative and sum to one are imposed to ensure the model is a well-defined convex combination of the component fields. However, while these conditions are sufficient for the model's theoretical definition, they do not guarantee parameter identifiability in a practical estimation setting. For that purpose, additional restrictions, such as imposing an ordering on the weights, would be required. From a modeling perspective, the key advantage of this framework is its flexibility.
By allowing for multiple transformations $\mathbf{M}_i$, the model gains flexibility in modeling various directional and scaling influences, making it suitable for applications where the propagation dynamics are heterogeneous or subject to multiple driving factors.
In particular, each component of \eqref{eq:DistributedFF_combination} is the solution of
\begin{align*}
\begin{cases}
     \frac{\partial Z_i(\bs,t)}{\partial t} + (\mathbf{M}_i\bv) \cdot \nabla Z_i(\bs,t) = 0,\\
     Z_i(\bs,0) = X_S(\bs).
\end{cases}
\end{align*}
In order to introduce a realistic appearance to $Z(\bs,t)$, we suggest selecting the transformations $\mathbf{M}_1, \dots, \mathbf{M}_{n_{\text{v}}}$ in \eqref{eq:DistributedFF_combination} randomly, i.e., drawing the rotation and scaling transformations from specified probability distributions. Moreover, for each transformation $\mathbf{M}_i$, the weight $p(\mathbf{M}_i\mathbf{v})$ determines the prominence of the corresponding geometric feature in the spatio-temporal phenomenon.

Notice that considering a random velocity in equation \eqref{eq:frozen_field} is an existing extension to the FF model. For instance, \cite{schlather2010some} considers the Cox-Isham rainfall model \citep{cox1988} where $\mathbf{v} \sim \mathcal{N}(\boldsymbol{\mu}, D)$ is drawn once for the entire field. In that framework, the mean velocity $\boldsymbol{\mu} \in \mathbb{R}^2$ determines the dominant advection direction, while the covariance matrix $D \in \mathbb{R}^{2 \times 2}$ governs the spread of velocities around this mean. The Distributed FF provides a discrete mixture generalization of this construction. 
To simplify the notation, let $\mathbf{v}_i = \mathbf{M}_i\mathbf{v}$ and $p_i = p(\mathbf{v}_i)$ for $i=1,\dots,n_v$.
Then, the empirical mean and covariance of the velocities $\{\mathbf{v}_i\}_{i=1}^{n_v}$ with weights $\{p_i\}_{i=1}^{n_v}$ are
\begin{equation*}
\bar{\mathbf{v}} = \sum_{i=1}^{n_v} p_i \mathbf{v}_i, \qquad \mathbf{C}_v = \sum_{i=1}^{n_v} p_i (\mathbf{v}_i - \bar{\mathbf{v}})(\mathbf{v}_i - \bar{\mathbf{v}})^\top,
\end{equation*}
which play the roles of $\boldsymbol{\mu}$ and $\mathbf{D}$ in the Gaussian formulation of Schlather (2010), respectively.
This discrete mixture representation offers greater modeling flexibility, as it allows for arbitrary velocity distributions through the choice of transformations $\{\mathbf{M}_i\}_{i=1}^{n_v}$ and weights $\{p_i\}_{i=1}^{n_v}$.

By allowing for random geometric transformations of the 
velocity, the Distributed FF model enables multiple propagation directions and scales within the same 
field. This structure naturally diminishes temporal correlation, as the 
superposition of components propagating with different velocities disrupts the 
perfect dependence observed in the frozen field model.

\begin{proposition}
Assume that $Z(\bs,t)$ is a distributed frozen field as specified in Definition \ref{def:distrFF}. Then, for fixed $(\bs, t) \in \mathbb{R}^2 \times \mathbb{R}^{+}$, the covariance function and the propagation paths of $Z(\bs,t)$ are given by
\begin{align}
    \label{eq:cov_distFF}
       \cov(Z(\bs,t), Z(\bs + \bh,t + \tau)) &= \sum_{i=1}^{n_{\text{v}}} p_i\sum_{j=1}^{n_{\text{v}}} p_j \, c_{XX}\left(\bh - \bv_j\tau - (\bv_j - \bv_i)t\right),
\end{align}
with propagation paths
    \begin{equation}
        \left(\bv_j\tau + (\bv_j - \bv_i)t, \tau\right), \quad i,j=1,\dots,n_{\text{v}},
\label{eq:prop_path_distrFF}
    \end{equation}
where $\bv_i = \mathbf{M}_i\mathbf{v}$ and $p_i = p(\mathbf{M}_i\mathbf{v})$ for $i=1,\dots,n_v$.
\end{proposition}

\begin{proof}
For fixed $(\bs, t) \in \mathbb{R}^2 \times \mathbb{R}^{+}$, the covariance of $Z(\bs,t)$ and $Z(\bs + \bh,t + \tau)$ is computed using the bilinearity of covariance:
\begin{align*}
\nonumber
&\cov(Z(\bs,t), Z(\bs + \bh,t + \tau)) \\
\nonumber
&= \cov\left(\sum_{i=1}^{n_{\text{v}}} p_i X_S \left(\bs - \bv_it\right), \sum_{j=1}^{n_{\text{v}}} p_j X_S \left(\bs + \bh - \bv_j(t + \tau)\right) \right) \\
\nonumber
&= \sum_{i=1}^{n_{\text{v}}} \sum_{j=1}^{n_{\text{v}}} p_i p_j \, \cov\left(X_S \left(\bs - \bv_it\right), X_S \left(\bs + \bh - \bv_j(t + \tau)\right)\right) \\
&= \sum_{i=1}^{n_{\text{v}}} \sum_{j=1}^{n_{\text{v}}} p_i p_j \, c_{XX}\left(\bh - \bv_j\tau - (\bv_j - \bv_i)t\right).
\end{align*}
Unity correlation occurs when the argument of $c_{XX}$ is zero, i.e.,
$$
\bh = \bv_j \tau + (\bv_j - \bv_i)t, \quad i,j=1,\dots,n_{\text{v}},
$$
thanks to the fact that $Z(\bs, t)$ is a convex combination of frozen fields.

\end{proof}

This proposition reveals how the Distributed FF model generalizes the classic FF by examining the structure of its covariance in \eqref{eq:cov_distFF}.
First, consider the terms where $i=j$. In this case, the expression simplifies to that of a standard FF: the time-dependent component $(\bv_j - \bv_i )t$ vanishes from the covariance argument, and the propagation path \eqref{eq:prop_path_distrFF} reduces to simple, straight-line advection, $\left( \bv_i \tau, \tau \right)$. These terms represent a weighted mixture of independent FFs, each with its own transformed velocity. Conversely, the terms where $i \neq j$ introduce novel cross-covariances between the different components. Specifically, $(\bv_j - \bv_i) t$ denotes the growing spatial displacement between two component paths over time $t$. The presence of this term makes the overall covariance function dependent on time, thus making the Distributed FF model \textit{non-stationary}. This complex behavior is known in physics as interference, referring to the interactions introduced by the superimposition of waves traveling in the same medium. Consequently, the distributed frozen field satisfies $\rho( \bv \tau, \tau) < 1$ unless $\bv_i = \mathbf{v}$ for all $i$, thereby preventing unrealistic perfect correlation along the propagation path of the FF.\\

In some settings it is more natural to assume the random field {\em locally} satisfies the wave equation. This means that, for every spatial location $\bs$
and time point $t$, we define a velocity $\mathbf{v}(\bs,t)$, where $\mathbf{v}: \mathbb{R}^2 \times \mathbb{R}^{+} \to \mathbb{R}^2$ is a continuous vector field. With this in mind, we propose the following stochastic process:

\begin{definition}[Evolving FF]
\label{def:evfrozen}
A random field $Z(\bs,t)$ is an {\em Evolving Frozen Field} if there is a deterministic velocity field $\mathbf{v}(\bs,t)\in{\mathbb{R}}^2$, and a random field $X_S(\bs)$ with $\bs\in {\mathbb{R}}^2$
and covariance function $c_{XX}(\bh)$ such that
\begin{equation}
Z(\bs,t)=X_S(\bs-\mathbf{v}(\bs,t)  t),
\label{eq:ev_frozen_field}
\end{equation}
with $t \in \mathbb{R}^+$.
\end{definition}

It is important to clarify the physical interpretation of \eqref{eq:ev_frozen_field}. In a general time-varying velocity field, the position of a particle is found by integrating the velocity field over its trajectory. Our definition provides a simplification of this picture, where the total displacement of the field at location $\bs$ and time $t$ is determined directly by the velocity vector at that final space-time coordinate. This makes the model analytically tractable while still capturing the essence of a locally varying advection.

The velocity field $\mathbf{v}(\bs,t)$ is an interpretable, infinite-dimensional parameter that connects our model to established concepts in fluid mechanics \citep{kundu2015fluid}. The Evolving FF is a flexible class of models that includes several well-known flows as special cases. For instance, if the velocity is constant in time, $\frac{\partial}{\partial t}\mathbf{v}(\bs,t)=\mathbf{0}$, this defines a {\em steady flow}. Examples of such steady flows include a {\em rigid body rotation} $\mathbf{v}(\bs)=(-k s_2, k s_1)$, a {\em stagnation point} $\mathbf{v}(\bs)=(k s_1, -k s_2)$, and a {\em vortex} $\mathbf{v}(\bs)=(-k s_2/s, k s_1/s)$ for $s=\sqrt{s_1^2+s_2^2}$. Other important special cases are defined by physical constraints, such as {\em incompressible flow} ($\nabla \cdot \mathbf{v}=\mathbf{0}$) or {\em irrotational flow} ($\nabla \times \mathbf{v}=\mathbf{0}$). Alternatively, a simpler non-stationary model assumes the velocity is constant in space, $\mathbf{v}(\bs,t)=\mathbf{v}(t)$.

Hence, model \eqref{eq:ev_frozen_field} is highly relevant, and determining its mathematical properties is intrinsically linked to the unobserved velocity field $\mathbf{v}(\bs,t)$. 
\begin{proposition}
Assume that $Z(\bs,t)$ is an evolving frozen field as specified by Definition~\ref{def:evfrozen}. For fixed $(\bs, t) \in \mathbb{R}^2 \times \mathbb{R}^{+}$, we have
\begin{align}
\label{eq:cov_approx_evolvingFF}
\nonumber
   &\cov(Z(\bs,t), Z(\bs + \bh,t + \tau)) =\\  &c_{XX}\left(\mathbf{h}-\mathbf{v}(\bs,t)\tau -\left\{  {\mathbf{D}}_\bs(\bs,t) \mathbf{h}+ D_t(\bs,t)\tau+O(\|\bh\|^2)+O(\tau^2)\right\}(t+\tau) \right),
\end{align}
with $ {\mathbf{D}}_\bs(\bs,t) = \begin{pmatrix}
\frac{\partial \text{v}_1(\bs,t)}{\partial s_1} & \frac{\partial \text{v}_1(\bs,t)}{\partial s_2}\\
\frac{\partial \text{v}_2(\bs,t)}{\partial s_1} & \frac{\partial \text{v}_2(\bs,t)}{\partial s_2}
\end{pmatrix}$ and $D_t(\bs,t) =  \left(\frac{\partial \text{v}_1(\bs,t)}{\partial t}, \frac{\partial \text{v}_2(\bs,t)}{\partial t}\right)^\top $. Moreover, the propagation path of $Z(\bs,t)$ is 
\begin{equation}
\label{eq:prop_path_evolvingFF}
\left((\mathbf{I}-{\mathbf{D}}_\bs(\bs,t)(t+\tau))^{-1}\left\{\mathbf{v}(\bs,t)+ D_t(\bs,t)(t+\tau)\right\}\tau,\tau\right),  \end{equation}
where $\mathbf{I}$ is the identity matrix of dimension $2$.
\end{proposition}
\begin{proof}
First, we apply local expansions on $\mathbf{v}(\bs,t)$. For fixed $(\bs, t) \in \mathbb{R}^2 \times \mathbb{R}^{+}$, we express the velocity field locally to be:
\begin{equation}
  \mathbf{v}(\bs + \mathbf{h},t + \tau) = \mathbf{v}(\bs,t) + {\mathbf{D}}_\bs(\bs,t) \mathbf{h}+ D_t \tau+O(\|\bh\|^2)+O(\tau^2). 
\label{eq:vel_approx}
\end{equation}
Then, we plug the expansion \eqref{eq:vel_approx} into the covariance of the evolving frozen field, i.e. 
\begin{align}
 \label{eq:cov_v(x,t)_approx}   
\nonumber
&\cov(Z(\bs,t), Z(\bs + \bh,t + \tau))=\\
\nonumber
&\cov\left( X(\bs-\mathbf{v}(\bs,t) t),X(\bs+\mathbf{h}-\mathbf{v}(\bs + \mathbf{h},t + \tau) (t+\tau))\right)  =\\
\nonumber
&   c_{XX}\left(\mathbf{h} - \left(\mathbf{v}(\bs + \mathbf{h},t + \tau) (t+\tau) -\mathbf{v}(\bs,t)t\right) \right)=\\
\nonumber
&c_{XX}\left(\mathbf{h} -\left\{\mathbf{v}(\bs,t) + {\mathbf{D}}_\bs(\bs,t) \mathbf{h}+ D_t(\bs,t)\tau+O(\|\bh\|^2)+O(\tau^2)\right\}(t+\tau) +\mathbf{v}(\bs,t)t\right)=\\
&c_{XX}\left(\mathbf{h}-\mathbf{v}(\bs,t)\tau -\left\{  {\mathbf{D}}_\bs(\bs,t) \mathbf{h}+ D_t(\bs,t)\tau+O(\|\bh\|^2)+O(\tau^2)\right\}(t+\tau) \right).
\end{align}
Finally, setting the argument of~\eqref{eq:cov_v(x,t)_approx} to zero yields the propagation path.
\end{proof}

Notice that the resulting covariance \eqref{eq:cov_approx_evolvingFF} and propagation path \eqref{eq:prop_path_evolvingFF} are local analytical approximations, valid for small spatial and temporal lags. \\

In summary, with the models presented in this section, we avoid having perfect correlation along $(\mathbf{v} \tau,\tau)$ in two ways: either by having random rotations and scaling of the constant velocity using transformations \eqref{eq:matrix_transformation}, or by ``perturbing" the propagation path as in \eqref{eq:prop_path_evolvingFF} by adding higher-order terms.

\subsection{Simulated Examples of a Cyclone}

To illustrate the behavior of these velocity-based generalizations, we now present simulation examples for both the Distributed and Evolving FF models. We choose to simulate the evolution of a tropical cyclone \citep{ dvorak1975tropical}, which has attracted a lot of interest in the climatology community for parametric models used in estimation (see for instance \cite{olander2007}, \cite{knaff2010}, \cite{pradhan2018}, \cite{maskey2020}). Our aim is to show that our proposed models can simulate complex dynamics that go well beyond those allowed by the FF model, and that are of interest in the climate science community.

Before discussing the simulated examples, we illustrate how to simulate from the two models introduced in Section \ref{sec:Distr_Ev_FF_framework}. First, the Distributed FF consists in a convex combination of $n_v$ realizations of FFs, each with their specified velocity and wight. Thus, to simulate a Distributed FF one just simulates each FF $n_v$ times with Algorithm \ref{alg:sim_FF}. Moreover, the simulation of an Evolving FF can also be carried out with a modification of Algorithm \ref{alg:sim_FF}. Specifically, when the space random field $X_S(\bs)$ is moved pointwisely in time, instead of shifting it according to a constant velocity, it is moved according to a velocity field $\bv(\bs, t)$. As a continuous function of space and time, $\bv(\bs, t)$ is discretized over the space and time grids. For practical implementation, we suggest discretizing $\bv(\bs, t)$ on the same space-time grid as the simulation. When it comes to establishing how large grid $\mathcal{G}^{*}$, should be, one can adopt $\text{v}_{\text{max}} = \max \{\lVert \text{v}_1 \rVert_\infty, \lVert \text{v}_2 \rVert_\infty\}$ for varying velocity $\bv(\bs,t) = (\text{v}_1(\bs,t), \text{v}_2(\bs,t))^\top $, where we determine $\lVert \text{v}_i \rVert_\infty = \inf \{ V \geq 0 : |\text{v}_i | \leq V \text{ for a.e. } (\bs,t) \}$, $i=1,2$. We include both algorithms in the supplementary materials.\\

We now illustrate the Distributed FF concept with a simulation where the component velocities are resampled at each time step, allowing the overall pattern to evolve. In particular, for a given time $t \in \{0,\dots,T\}$, we are interested in combining rotation and translation of the random field. We first draw a set of $n_\theta$ random angles,
\begin{equation}
\label{eq:random_rotations_t}
    \theta_{1,t},\dots,\theta_{n_\theta, t} \stackrel{iid}{\sim} f_{R}(\theta;\boldsymbol{\alpha}_R),
\end{equation}
and a single random velocity vector,
\begin{equation}
\label{eq:random_translation_t}
  \bv_t \sim f_{T}(\bv;\boldsymbol{\alpha}_T).
\end{equation}
Equation \eqref{eq:random_rotations_t} denotes the $n_\theta$ angles that define the random rotations of the field, where $f_R$ is a probability density function (PDF). Equation \eqref{eq:random_translation_t} describes the velocity $\bv_t$ with which the entire field is advected at time $t$.

For a fixed $t$, we obtain the distributed frozen field as a weighted average of multiple rotated and translated fields:
\begin{align}
\label{eq:DistributedFF_combination_discrete_rotation}
Z(\bs,t)&=\sum_{i=1}^{n_\theta} p(\theta_{i,t}) X^{\theta_{i,t}}_S(\bs-\bv_t t).
\end{align}
Here, $X^{\theta_{i,t}}_S$ denotes the base random field $X_S$ after being rotated by the angle $\theta_{i,t}$ around a fixed center of rotation, $C$. The term $(\bs-\bv_t t)$ indicates that this rotated field is evaluated at a coordinate shifted by the velocity vector, effectively modeling advection.
The weights $p(\theta_{i,t})$ are constructed to form a convex combination, ensuring that $p(\theta_{i,t})\ge 0$ and $\sum_{i=1}^{n_\theta} p(\theta_{i,t}) = 1$. They are derived from the PDF, $f_R$, evaluated at each sampled angle. Specifically, the weight for each rotation is calculated by normalizing the PDF value at that angle:
\begin{align*}
p(\theta_{i,t}) = \frac{f_R(\theta_{i,t}; \boldsymbol{\alpha}_R)}{\sum_{j=1}^{n_\theta} f_R(\theta_{j,t}; \boldsymbol{\alpha}_R)}.
\end{align*}
This approach ensures that rotations corresponding to angles with higher probability density are given more weight in the final mixture. 
Notice that the Distributed FF with time-varying transformations admits a natural interpretation as an expected field. Specifically, the discrete sum \eqref{eq:DistributedFF_combination_discrete_rotation} can be viewed as a Monte Carlo approximation to the expectation with respect to the rotation angle and velocity, i.e., $\int_{\mathbb{R}^2} \int_{-\pi}^{\pi} X_S^{\theta}(\mathbf{s} - \mathbf{v} t) \, f_R(\theta; \boldsymbol{\alpha}_R) \, f_T(\mathbf{v}; \boldsymbol{\alpha}_T) \, d\theta \, d\mathbf{v}$. This connects directly to the framework of \cite{schlather2010some}, who considered the expected covariance structure arising from a Gaussian random velocity. Our formulation extends this by incorporating both random rotations and translations, with the discrete representation allowing for non-Gaussian distributions and improving in accuracy as $n_\theta$ grows.

\begin{figure}[htb!]
\begin{center}
\includegraphics[width=\textwidth]{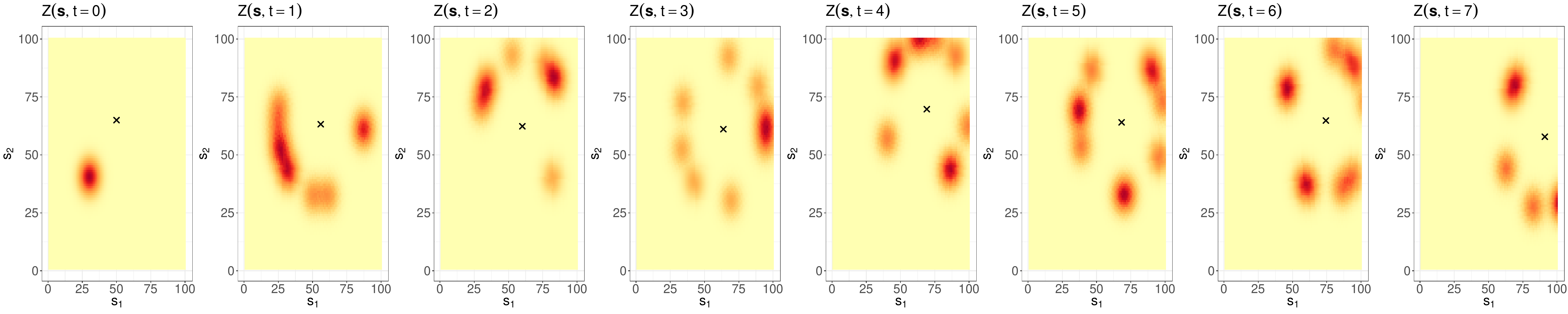}
\end{center}
\caption{Distributed FF simulation of a cyclone.
}
\label{fig:Dist_FF_vst_sim}
\end{figure}

To simulate the example in Figure \ref{fig:Dist_FF_vst_sim}, we set $n_\theta=10$. For each $t \in \{0,\dots,7\}$, the angles are drawn from a Wrapped Normal distribution, $\theta_{i,t} \stackrel{iid}{\sim} \mathcal{WN}(0, 5)$, and the velocity is drawn from a multivariate Normal distribution, $\bv_t\sim \mathcal{N}(\boldsymbol{\mu}, \mathbf{I})$, with mean $\boldsymbol{\mu} = (5,0)^\top $ and the identity as covaraince matrix. The rotation is performed around a fixed center $C=(50, 65)$. The \textit{x} marks in the plot indicate the position of this rotation center relative to the advected field, given by $C_t = C + \bv_t t$. For this simulation, the initial field at $t=0$, $Z(\boldsymbol{s}, 0) = X(\boldsymbol{s})$, is a fixed, deterministic image rather than a realization of a random field. We deliberately choose a simple shape, a faded circle on a constant background, to provide a clear visual benchmark. This approach allows the model's intrinsic dynamics to be observed directly and clearly.

\begin{figure}[htb!]
\begin{center}
\includegraphics[width=\textwidth]{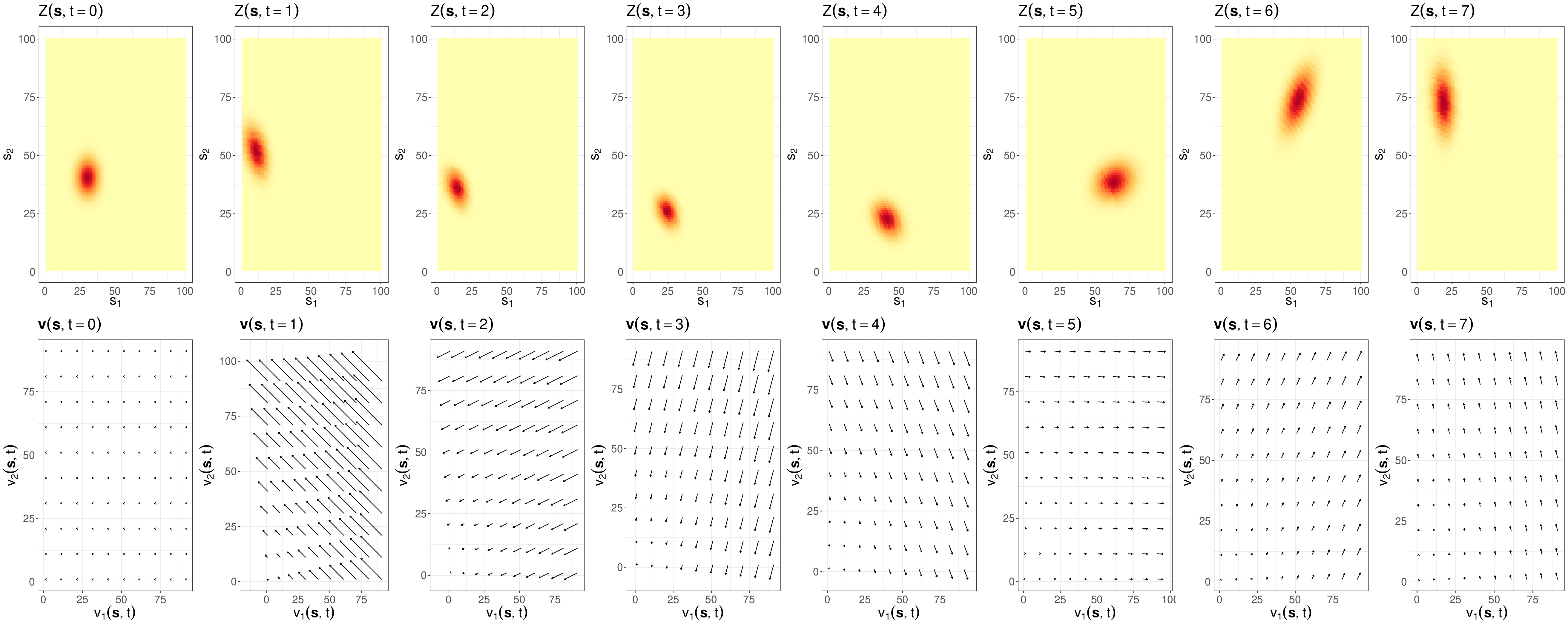}
\end{center}
\caption{Evolving FF simulation of a cyclone (top row) and the corresponding velocity field} $\bv(\bs,t)$ at every time point (bottom row).
\label{fig:het_FF_vst_sim_square}
\end{figure}

It could be of interest to simulate how the cyclone evolves locally, namely how a single portion of it travels and gets deformed along time. We simulate such trajectory with an Evolving FF model. In particular, Figure \ref{fig:het_FF_vst_sim_square} shows
an example in which velocity field $\bv(\bs, t)$ is such that
\begin{equation}
\label{eq:ex_vst}
    \bv (\bs,t) = \left ( \frac{R(\bs) \cdot \text{cos}( t)}{t+1},\frac{ R(\bs)\cdot\text{sin}(t)}{t+1} \right )^\top ,
\end{equation}
with
\begin{align*}
     R(\bs)= \frac{|| \bs ||}{2}.
\end{align*}
Velocity field \eqref{eq:ex_vst} describes a spiral whose radius $R(\bs)$ changes at each spatial location. Hence, the magnitude of the point-wise elements of the velocity field depends on both the location and the time, while their direction only on the latter. In the top and bottom rows of Figure \ref{fig:het_FF_vst_sim_square} we show the traveling field and the corresponding velocity fields changing over time, respectively. Notice that we used the same initial image as in Figure~\ref{fig:Dist_FF_vst_sim}, and that the velocity fields have been re-scaled for representation purposes. As we would expect in the observed temporal evolution of a tropical cyclone, the points closest to the center, i.e. the origin, get less distorted than those further away from it.

\subsection{Comparison with Numerical Solvers}
We now compare our simulation methods with numerical solutions of the corresponding PDEs. A comparison with numerical PDE solvers is particularly appropriate, as both our method and these solvers are fundamentally constructed upon a specified advection velocity. While PDE solvers do not constitute ground truth, they provide a deterministic reference that targets the same dynamics from a different computational angle. Since the FF satisfies the transport equation \eqref{eq:transport_PDE}, and the Distributed FF satisfies an advection-diffusion equation under appropriate closure assumptions as we show later in this section, the comparison serves to verify that our stochastic constructions recover the expected behavior.

We consider simulations on a spatial grid $\mathcal{G} = [0,100] \times [0,100]$ and time grid $t \in \{0, 1, 2,3\}$. We use package \texttt{ReacTran} \citep{reactran} to compute the solution of the PDEs. Moreover, to compare the simulated space-time random field $Z(\bs,t)$ and the solution of the numerical solver $Z_{\text{PDE}}(\bs,t)$, we use Root Mean Square Error (RMSE)
\begin{align*}
    \text{RMSE}(t) = \sqrt{\frac{1}{n^2} \sum_{\bs \in \mathcal{G}} \left(Z(\bs, t) - Z_{\text{PDE}}(\bs, t)\right)^2}, \quad t\in \{0, 1, 2,3\}.
\end{align*}

\begin{figure}[htb!]
\begin{center}
\includegraphics[width=\textwidth]{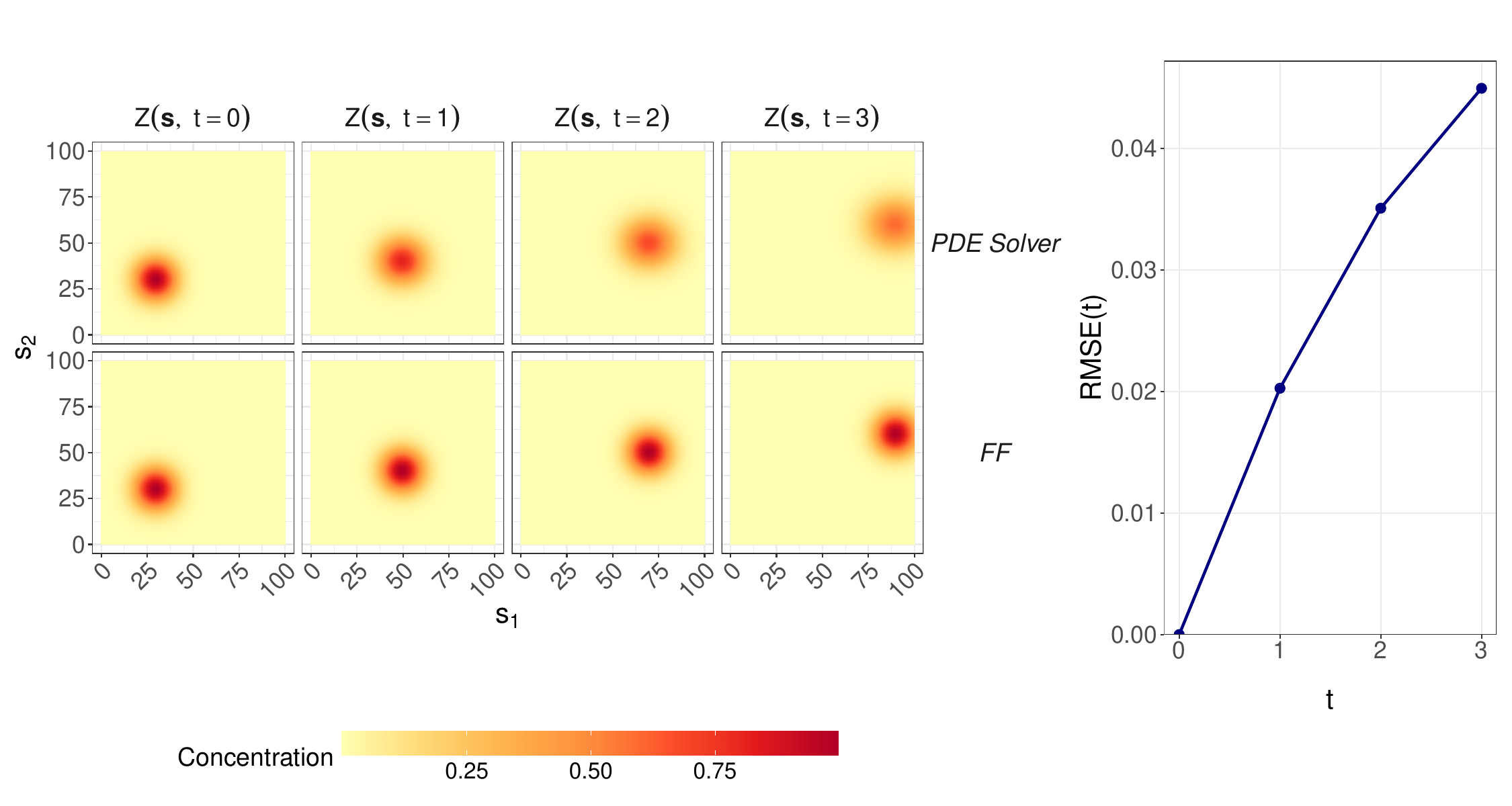}
\end{center}
\caption{On the left, solution of the transport PDE (top row) and the corresponding simulated Frozen Field (bottom row). On the right, the RMSE between two images at each time point.
}
\label{fig:FF_transport}
\end{figure}

As mentioned in Section~\ref{sec:background_theory}, the FF corresponds to the solution of the pure transport PDE \eqref{eq:transport_PDE}. Thus, the most natural first comparison we make is between our simulated FF and the solution of a transport equation. We simulate a FF with $\bv=(20, 10)^\top$, and we solve \eqref{eq:transport_PDE} with a no-flux boundary condition. 
The results are shown in Figure \ref{fig:FF_transport}. Our proposed simulation method closely resembles the numerical solution of the PDE. However, the latter exhibits numerical diffusion, a well-known artifact of finite-difference schemes for hyperbolic equations \citep{LeVeque_2002}, causing the concentration of the image to dissipate slightly over time. This dissipation is quantified in the RMSE plot. The error is zero at $t=0$, as both methods are initialized with the identical field, but grows over time as numerical diffusion accumulates in the PDE solver. By construction, our method samples the advected field directly and does not suffer from this artifact, illustrating a practical advantage for simulating pure transport.

We now turn to the comparison of a Distributed FF with the solution of a transport-diffusion PDE. We show how the superposition of multiple FFs, each with a distinct advection velocity, gives rise to both a mean advective transport and an effective diffusive term.
We begin by considering that the field is a superposition of multiple components, each transported by a distinct advection velocity $\mathbf{v}_i$ with probability $p_i$, as introduced in Definition~\ref{def:distrFF}. Recall from Section~\ref{sec:Distr_Ev_FF_framework} the mean velocity $\overline{\bv} = \sum_{i=1}^{n_v} p_i \mathbf{v}_i$ and velocity covariance $\mathbf{C}_{\mathbf{v}} = \sum_{i=1}^{n_v} p_i (\mathbf{v}_i - \overline{\bv})(\mathbf{v}_i - \overline{\bv})^\top$. We define the velocity deviations $\Delta \bv_i = \bv_i - \overline{\bv}$. Similarly to \cite{christakos2017spatiotemporal}, treating the velocity as the sum of a mean and fluctuations is an advantageous framework, as we shall see next.

It can be shown (see supplementary materials) that for fixed $(\bs,t)$, the governing equation for the DFF field $Z(\bs,t)$ is:
\begin{equation}
\label{eq:DistributedFF_diffusion}
\frac{\partial Z(\bs, t)}{\partial t} + \overline{\bv} \cdot \nabla Z(\bs, t) = - \nabla \cdot \left( \sum_{i=1}^{n_v} p_i\, Z_i(\bs, t) \Delta \bv_i \right).
\end{equation}
The left-hand side of \eqref{eq:DistributedFF_diffusion} represents the pure advection of the field $Z(\bs,t)$ at the mean velocity $\overline{\bv}$. The term on the right is the divergence of the ``fluctuating flux", which arises from the velocity deviations. The equation in its current form is not closed, as the flux depends on the individual components $Z_i(\bs,t)$. To create a macroscopic equation for $Z(\bs,t)$, we must impose further assumptions.

We adopt the widely used gradient-diffusion hypothesis, which assumes that the fluctuating flux is proportional to the negative gradient of the mean field. In other words, this implies that the random motion of particles will naturally cause them to move from a more crowded region to a less crowded one.
In our model, this entails
\begin{align}
\label{eq:fickian_closure}
\sum_{i=1}^{n_v} p_i \, Z_i(\bs, t) \Delta \bv_i = -\mathbf{C}\nabla Z(\bs,t),
\end{align}
where $\mathbf{C}$ is the diffusion tensor. Substituting \eqref{eq:fickian_closure} into \eqref{eq:DistributedFF_diffusion}, we arrive at the advection-diffusion equation:
\begin{align*}
\frac{\partial Z(\bs, t)}{\partial t} + \overline{\bv} \cdot \nabla Z(\bs, t) = \nabla \cdot (\mathbf{C}\nabla Z(\bs,t)).
\end{align*}
This derivation provides a formal link between our Distributed FF model and the advection-diffusion PDE. The advection-diffusion equation derived here can be compared to the 
spatio-temporal SPDE framework presented by \cite{carlotto2024}. 
Their model includes a random forcing term on the right-hand side that 
introduces stochastic variability through the product of temporal white 
noise and spatial noise. 
Our Distributed FF formulation differs by having zero on the 
right-hand side, representing deterministic transport-diffusion dynamics 
where randomness originates entirely from the distribution of velocities 
rather than from external forcing. These represent two distinct mechanisms 
for generating spatio-temporal variability, where our approach models uncertainty 
in the advective flow itself, while their framework adds fluctuations to 
otherwise deterministic dynamics. In physical systems, both random advection by varying currents and residual turbulent noise may be 
present, suggesting these formulations address complementary aspects of 
spatio-temporal phenomena.

The identification of the diffusion tensor $\mathbf{C}$ with the covariance of the velocity distribution \citep{Taylor1954} has a physical interpretation. Specifically, diffusion is a macroscopic process of spatial spreading. In the Distributed FF model, the microscopic cause of this spreading is the fact that different additive components of the field are being transported at different velocities. The velocity covariance matrix, namely $\mathbf{C}_{\mathbf{v}}$, is the precise mathematical tool that quantifies the magnitude and orientation of this velocity spread. Therefore, the diffusion tensor $\mathbf{C}$ in the macroscopic PDE is the direct physical manifestation of the velocity covariance, and we can assume that $\mathbf{C}$ is proportional to $\mathbf{C}_{\mathbf{v}}$. This principle allows us to construct a Distributed FF that mimics a desired advection-diffusion process by defining a velocity distribution with the appropriate mean and covariance.

\begin{figure}[htb!]
\begin{center}
\includegraphics[width=\textwidth]{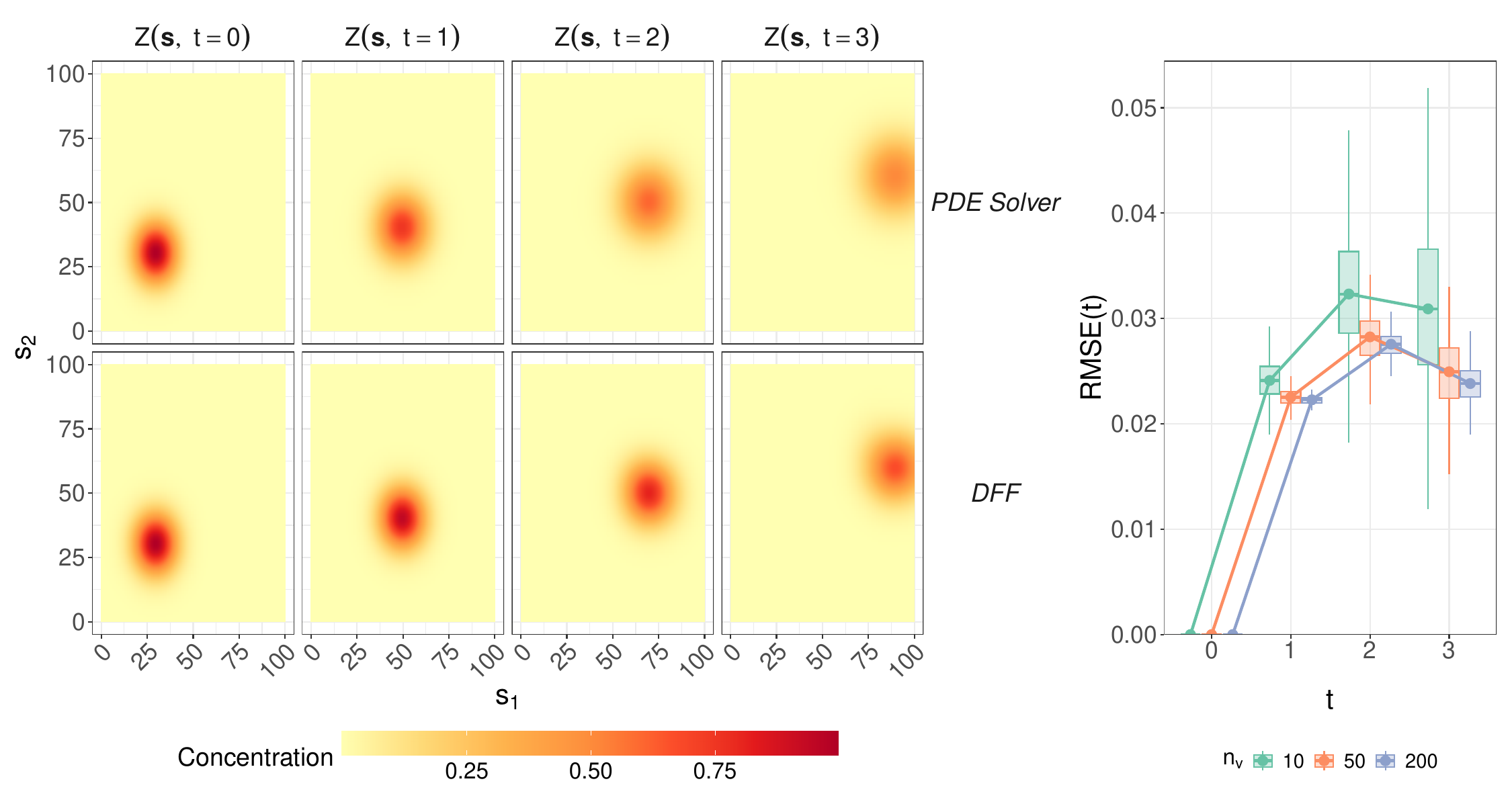}
\end{center}
\caption{On the left, solution of the transport-diffusion PDE (top row) and a simulated Distributed Frozen Field with $n_v=50$. (bottom row). On the right, the boxplots of RMSE between two images at each time point across $500$ iterations and for $n_v=10,50,200$.
}
\label{fig:DFF_diffution}
\end{figure}

To validate our theoretical framework, we compare our Distributed FF model against a numerical solution of the advection-diffusion equation. 
The PDE was solved on the same spatial and time grids as in the previous FF example, with a velocity of $\overline{\bv} = (20, 10)^\top $ and a diffusion tensor of $\mathbf{C} = \begin{pmatrix}
    5 & 0\\
    0 & 1.5
\end{pmatrix}$.
The Distributed FF model was constructed by drawing $\{\mathbf{v}_i\}_{i=1}^{n_v}$ from a multivariate normal distribution whose parameters were set to match the PDE, i.e. $\bv_i \stackrel{iid}{\sim} \mathcal{N}(\overline{\bv}, \mathbf{C})$. The weights used are $p_i = 1/n_v$, for each $i=1,\dots,n_v$. 
In Figure \ref{fig:DFF_diffution} we compare a single Distributed FF realization with $n_v = 50$ against the PDE numerical solution, showing qualitative agreement between the two. This agreement is notable because the two methods approach the same advection-diffusion dynamics through fundamentally different mechanisms: the PDE solver discretizes the Laplacian operator directly, while the Distributed FF constructs diffusive behavior from the superposition of fields advected at different velocities.

To quantify the model's convergence and performance, we carry out a simulation study of $500$ iterations using different numbers of components, $n_v = 10, 50, 200$. The right panel of Figure~\ref{fig:DFF_diffution} quantifies the model's performance against the PDE solution. The results clearly indicate that as $n_v$ increases, both the median RMSE and its variance decrease across all time points, confirming our theoretical derivation and demonstrating that the approximation error is controllable. The shape of the RMSE curves also provides insight into the process. As established in the pure transport case, the error is zero at $t=0$ because both methods are initialized with the identical field. The RMSE then rises as the approximation error accumulates, particularly during the initial phase where the concentration gradients are steepest. Subsequently, the curve plateaus because the diffusion process itself smooths the field over time. As the gradients lessen, the rate at which new error is generated decreases, causing the total error to stabilize.

\subsection{Application to Precipitation Advection: Hurricane Florence}

To illustrate the practical relevance of the proposed framework, we apply the FF and Evolving FF models to reproduce the rainfall evolution of during Hurricane Florence, a slow-moving tropical cyclone that affected the Carolinas in mid-September 2018. Making landfall near Wrightsville Beach, North Carolina, Florence was characterized by a collapse in environmental steering currents that reduced its forward velocity to approximately 9.3 km/h. This stalling behavior led to persistent rainbands over the region, resulting in catastrophic freshwater flooding and record-breaking rainfall accumulations \citep{stewart2019florence}.

We utilize the NCEP Stage IV precipitation product \citep{nelson2016assessment}, a multi-sensor analysis combining radar and gauge observations provided by the National Centers for Environmental Prediction, as source of data. This dataset, available via the \texttt{R} package \texttt{stars} \citep{pebesma2023spatial, stars}, provides hourly rain accumulations (kg/m$^2$) at approximately 4 km spatial resolution. The analysis focuses on a 3-hour window from 14:00 to 17:00 UTC on September 14th 2018, immediately following landfall.

We define the spatial grid $\mathcal{G} = [1, 87] \times [1, 118]$ as an $87 \times 118$ pixel lattice, covering an area of approximately 350 km $\times$ 470 km centered over southeastern North Carolina.
The temporal domain is discretized into 15-minute intervals, with intermediate frames obtained by linear interpolation between consecutive hourly observations. Let $t \in \{0, \dots, 12\}$ denote the time steps, where $t=0$ corresponds to the initialization time at 14:00 UTC, and $t=12$ corresponds to the final evaluation time at 17:00 UTC. The velocity field $\mathbf{v}(\mathbf{s},t)$ is estimated via optical flow using a local block-matching algorithm \citep{horn1981, bowler2004}. For each location $(\mathbf{s}, t)$, the local velocity vector is determined by finding the displacement maximizing the cross-correlation between consecutive frames within a local neighborhood $\mathcal{B}_{\mathbf{s}}$. Specifically, using a block parameter $b=9$ pixels, this neighborhood is defined as $\mathcal{B}_{\mathbf{s}} = \left\{ \mathbf{u} \in \mathcal{D} : \| \mathbf{u} - \mathbf{s} \|_\infty \leq 4 \right\}$. The estimator is given by:
\begin{align*}
\widehat{\mathbf{v}}(\mathbf{s}, t) = \arg\max_{(\Delta s_1, \Delta s_2)} \, \text{Corr}\left( Z(\mathcal{B}_{\mathbf{s}}, t), \, Z(\mathcal{B}_{\mathbf{s}} + (\Delta s_1, \Delta s_2), t + 1) \right),
\end{align*}
where $(\Delta s_1, \Delta s_2) \in \{-4, \ldots, 4\}^2$ represent candidate spatial displacements in grid units. To avoid spurious matches in regions of uniform intensity, we assign zero velocity where the standard deviation of $Z(\mathcal{B}_{\mathbf{s}}, t)$ is below 0.2 kg/m$^2$, and we retain only displacements achieving a correlation above 0.4. The resulting field is smoothed using a Gaussian kernel with $\sigma=2$ pixels.

For the Evolving FF, the spatially-varying field $\widehat{\mathbf{v}}(\mathbf{s}, t)$ is used directly at each time step. Notice that, prior to velocity estimation, each frame is smoothed using a Gaussian kernel with $\sigma=1$ pixel to reduce noise. For the standard FF, we compute a single constant velocity by spatially averaging the initial velocity field:
\begin{align*}
\widehat{\overline{\mathbf{v}}} = \frac{1}{|\mathcal{G}|} \sum_{\mathbf{s} \in \mathcal{G}} \widehat{\mathbf{v}}(\mathbf{s}, t=0).
\end{align*}

Forecasts are generated by advecting the precipitation field according to the estimated velocities, using bilinear interpolation for sub-pixel displacements. 
All metrics are computed on the interior of the domain, excluding a 15-pixel buffer from the boundaries to avoid artifacts. This results in an evaluation region of size $57 \times 88$ pixels.
Performance is quantified using RMSE:
\begin{align*}
\text{RMSE}(t) = \sqrt{ \frac{1}{|\mathcal{G}|} \sum_{\mathbf{s} \in \mathcal{G}} \left( Z(\mathbf{s},t) - \widehat{Z}(\mathbf{s},t) \right)^2 },
\end{align*}
and the spatial correlation between the observed field $Z(\mathbf{s},t)$ and the simulated field $\widehat{Z}(\mathbf{s},t)$:
\begin{align*}
\text{Corr}(t) = \frac{ \sum_{\mathbf{s} \in \mathcal{G}} \left(Z(\mathbf{s},t) - \overline{Z}(t)\right) \left(\widehat{Z}(\mathbf{s},t) - \overline{\widehat{Z}}(t) \right) }{ \sqrt{ \sum_{\mathbf{s} \in \mathcal{G}} \left( Z(\mathbf{s},t) - \overline{Z}(t) \right)^2 } \, \sqrt{ \sum_{\mathbf{s} \in \mathcal{G}} \left(\widehat{Z}(\mathbf{s},t) - \overline{\widehat{Z}}(t) \right)^2 }},
\end{align*}
where $\overline{Z}(t) = \frac{1}{|\mathcal{G}|} \sum_{\mathbf{s} \in \mathcal{G}} Z(\mathbf{s},t) $ and $\overline{\widehat{Z}}(t) = \frac{1}{|\mathcal{G}|} \sum_{\mathbf{s} \in \mathcal{G}} \widehat{Z}(\mathbf{s},t)$. 

The estimated constant velocity for the FF was $\widehat{\overline{\mathbf{v}}} = (-0.21, -0.08)^\top$ grid cells per 15-minute step. This corresponds to a translation speed of approximately 3.6 km/h (0.89 grid cells per hour) toward the south-southwest ($202^\circ$). This estimate is notably lower than the overall storm motion of 9.3 km/h reported by \cite{stewart2019florence}, reflecting the fact that the optical flow captures the slow rotation of spiral bands during the stalling phase rather than just the translation of the eye. Over the 3-hour forecast horizon, this yields a net displacement of roughly 11 km, corresponding to less than three grid cells. 

\begin{figure}[htb!]
    \centering
    \includegraphics[width=\linewidth]{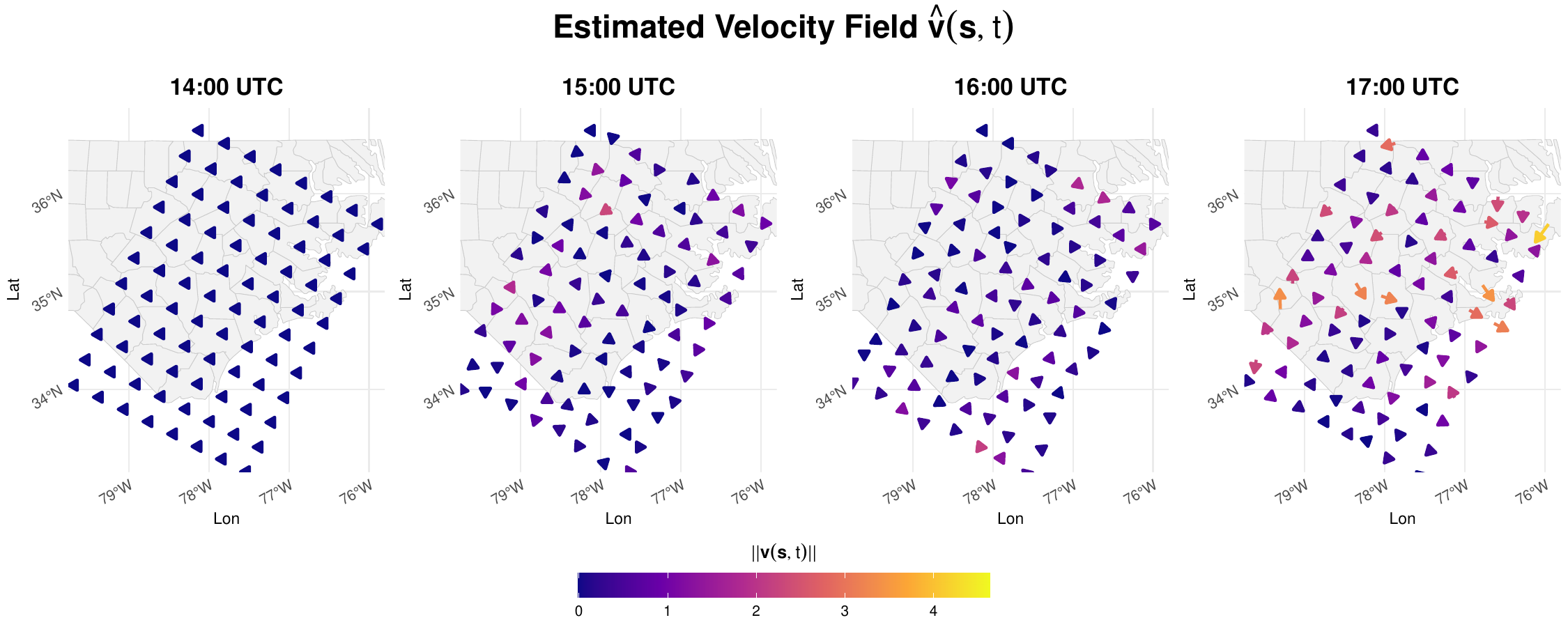}
    \caption{Estimated velocity field $\widehat{\mathbf{v}}(\mathbf{s},t)$ on September 14, 2018, at four consecutive hours between 14:00 UTC and 17:00 UTC. The arrows indicate flow direction, while the color scale represents the velocity magnitude $||\mathbf{v}(\mathbf{s},t)||$.}
    \label{fig:vel_field}
\end{figure}

The advantage of the Evolving FF is evident in Figure~\ref{fig:vel_field}, which displays the estimated velocity field $\widehat{\mathbf{v}}(\mathbf{s},t)$. Unlike the single averaged vector of the Frozen FF, the evolving field captures substantial spatial heterogeneity. The vector plot reveals the presence of rotation of the system, and identifies localized regions of rapid advection associated with convective rainbands.

\begin{figure}[htb!]
    \centering
    \includegraphics[width=\linewidth]{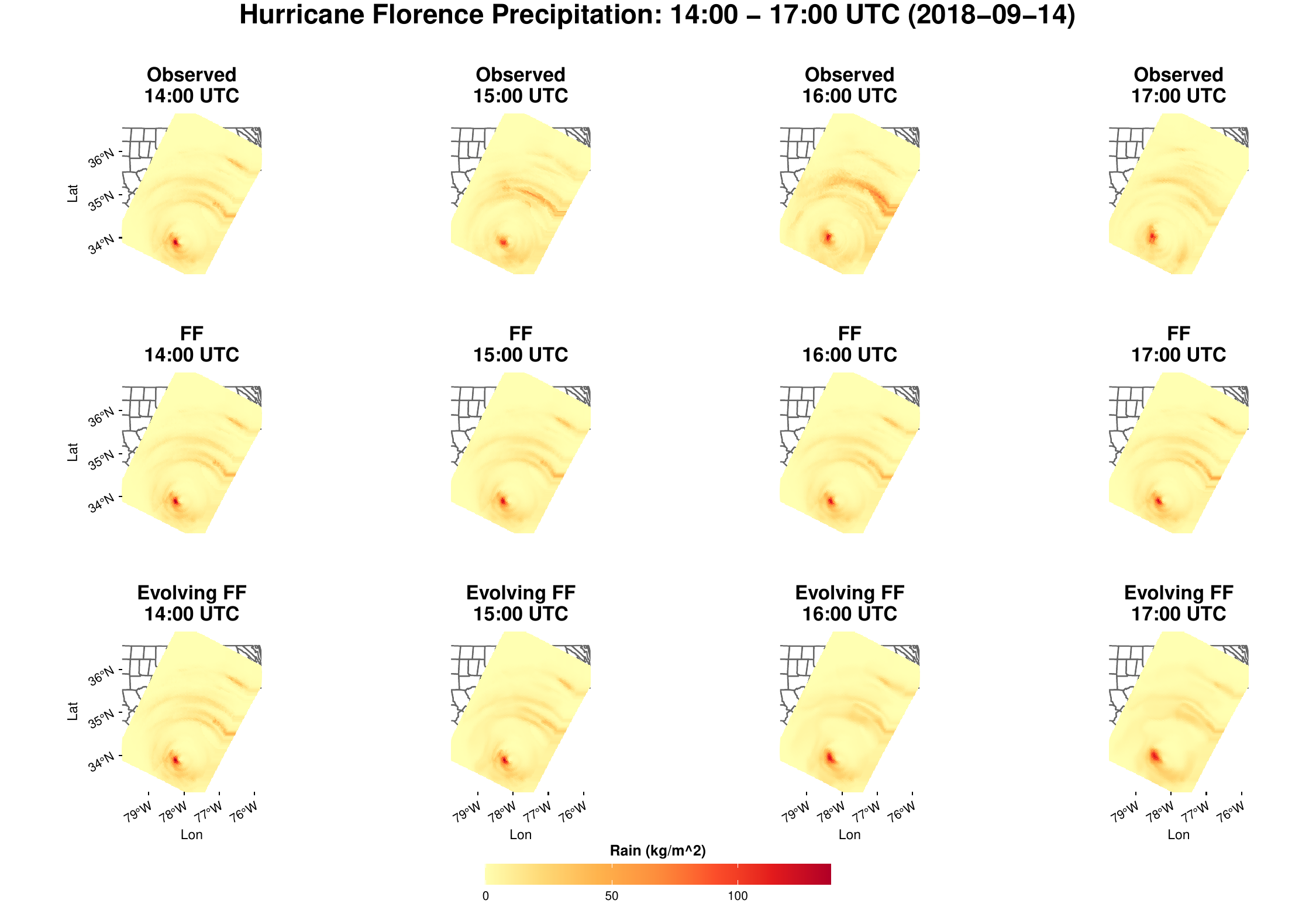}
    \caption{Observed random field (first row), simulated FF (second row), and simulated Evolving FF (third row), at four consecutive hours between 14:00 UTC and 17:00 UTC, on September 14, 2018.}
    \label{fig:simulated_fields}
\end{figure}

Figure~\ref{fig:simulated_fields} shows the observed and forecast precipitation fields at hourly intervals: $t=0, 4, 8, 12$, corresponding to 14:00, 15:00, 16:00, and 17:00 UTC. While visual differences are subtle at early lead times, divergence becomes apparent by 17:00 UTC. 
The FF simulation maintains a pattern closely tied to the initialization, whereas the Evolving FF better captures the evolution of precipitation.

Finally, Table~\ref{tab:metrics} summarizes the quantitative comparison. The Evolving FF consistently outperforms the FF. For instance, at $\Delta t = 180$ minutes, the Evolving FF achieves a correlation of 0.76 compared to 0.48 for the FF, and reduces RMSE from 8.7 to 6.4 kg/m$^2$. 

We conclude that the advantage of the Evolving FF derives from its ability to accommodate local variations in the flow, which a rigid advection model cannot represent.

\begin{table}[htb!]
\centering
\caption{Performance metrics for Hurricane Florence precipitation forecasting on September 14th, 2018. Lead time $\Delta t$ is measured relative to the initialization at 14:00 UTC.}
\label{tab:metrics}
\begin{tabular}{cc|cc|cc}
\hline
\multicolumn{2}{c|}{Time} & \multicolumn{2}{c|}{$\text{Corr}(t)$} & \multicolumn{2}{c}{RMSE (kg/m$^2$)} \\
$\Delta t$ & UTC & FF & Evolving FF & FF & Evolving FF \\
\hline
0 min   & 14:00 & 1.000 & 1.000 & 0.00 & 0.00 \\
60 min  & 15:00 & 0.777 & 0.832 & 6.03 & 5.26 \\
120 min & 16:00 & 0.487 & 0.628 & 10.78 & 9.48 \\
180 min & 17:00 & 0.480 & 0.761 & 8.72 & 6.39 \\
\hline
\end{tabular}
\end{table}

\section{Generalization via Relaxation of the Spectrum Shape}

In this section, we consider a different type of generalization of the FF model, namely by relaxing the shape of its spectrum \eqref{eqn:ff_spectrum}. We first propose a new process by defining its spectrum. Then, we analyze the role of the parameters in such definition with an example.

\subsection{Framework}

For a fixed wavenumber $\bk$ and velocity $\bv$, the spectrum of a FF \eqref{eqn:ff_spectrum} consists of the spectrum $S_{XX}(\bk)$ of the spatial random field multiplied by a Dirac delta centered in $\omega=-\bk^\top \mathbf{v}$. Such multiplicative structure characterizing $S_{ZZ}(\bk, \omega)$ makes it particularly convenient for us to modify the part of the spectrum that depends on velocity $\bv$. For instance, replacing the Dirac deltas with a ``milder'' singularity would ensure that the spectrum includes more frequencies neighbouring $\omega=-\bk^\top \mathbf{v}$. This introduces temporal decorrelation into the model, which gains the flexibility to represent more complex and realistic phenomena where features naturally change, dissipate, or evolve over time.

The following proposition provides the theoretical formalization for this intuition, proving that for a broad class of stationary fields, perfect temporal correlation is a unique property of the FF model.

\begin{proposition}
\label{prop:corr}
Let $Z(\bs, t)$ be a stationary space-time random field with a valid spectrum $S_{ZZ}(\mathbf{k}, \omega)$ and covariance function $c_{ZZ}(\mathbf{h}, \tau)$. Then, for any $\tau > 0$, the correlation along the path $(\mathbf{v}\tau, \tau)$ satisfies
$$
\rho(\mathbf{v} \tau, \tau) = \frac{|c_{ZZ}(\mathbf{v} \tau, \tau)|}{c_{ZZ}(0,0)} < 1,
$$
unless $Z(\bs, t)$ is a frozen field with velocity $\mathbf{v}$.
\end{proposition}
\begin{proof}
By definition, the covariance $c_{ZZ}(\mathbf{h}, \tau)$ is the inverse Fourier transform of the spectrum $S_{ZZ}(\mathbf{k}, \omega)$:
$$
 c_{ZZ}(\mathbf{h}, \tau) = \iiint S_{ZZ}(\mathbf{k}, \omega) \, e^{2\pi i (\mathbf{k}^\top \mathbf{h}  + \omega \tau)} \, \mathrm{d}\omega \, \mathrm{d}^2\mathbf{k}.
$$
Evaluating this along the path $(\mathbf{v}\tau, \tau)$ and at the origin gives:
\begin{align*}
    c_{ZZ}(\mathbf{v}\tau, \tau) &= \iiint S_{ZZ}(\mathbf{k}, \omega) \, e^{2\pi i \tau(\mathbf{k}^\top \mathbf{v}  + \omega)} \, \mathrm{d}\omega \, \mathrm{d}^2\mathbf{k}, \\
    c_{ZZ}(0, 0) &= \iiint S_{ZZ}(\mathbf{k}, \omega)  \, \mathrm{d}\omega \, \mathrm{d}^2\mathbf{k}.
\end{align*}
By the triangle inequality, we have:
$$
    \left| \iiint S_{ZZ}(\mathbf{k}, \omega) \, e^{2\pi i \tau(\mathbf{k}^\top \mathbf{v}  + \omega)} \, \mathrm{d}\omega \, \mathrm{d}^2\mathbf{k} \right| \leq \iiint S_{ZZ}(\mathbf{k}, \omega) \, \left|e^{2\pi i \tau(\mathbf{k}^\top \mathbf{v}  +\omega)} \right| \, \mathrm{d}\omega \, \mathrm{d}^2\mathbf{k}.
$$
Since $|e^{i\theta}| = 1$ for any real $\theta \in \mathbb{R}$, the inequality simplifies to $|c_{ZZ}(\mathbf{v}\tau, \tau)| \leq c_{ZZ}(0, 0)$.
Equality can only hold if the argument of the integral, $S_{ZZ}(\mathbf{k}, \omega) e^{2\pi i \tau(\mathbf{k}^\top \mathbf{v} + \omega)}$, has a constant phase for all $(\mathbf{k}, \omega)$ where the spectrum $S_{ZZ}$ is non-zero. Given that $S_{ZZ}(\mathbf{k}, \omega)$ is a real and non-negative function, it does not affect the phase. Therefore, the phase of the complex exponential term $e^{2\pi i \tau(\mathbf{k}^\top \mathbf{v} + \omega)}$ must be constant, which implies that its exponent, $2\pi \tau(\mathbf{k}^\top \mathbf{v} + \omega)$, must be constant. This condition must hold for all $(\mathbf{k}, \omega)$ with non-zero spectral energy. If we consider the point $(\mathbf{k}=\mathbf{0}, \omega=0)$, which must also satisfy this relation, we find that $2\pi\tau(0+0) = 0$.
Thus, with the constant being zero, the condition for equality becomes $\mathbf{k}^\top \mathbf{v} + \omega = 0$ on the set of wavenumber-frequency $\{(\bk,\omega) \in \mathbb{R}^2 \times \mathbb{R}:  S_{ZZ}(\mathbf{k}, \omega) > 0\}$, implying $\omega = -\mathbf{k}^\top\mathbf{v}$ in the same set. 
This means the spectrum $S_{ZZ}$ must be entirely supported on the hyperplane defined by $\omega = -\mathbf{k}^\top\mathbf{v}$, which is precisely the spectral definition of a FF, for which $\rho(\bv\tau,\tau)=1$.

\end{proof}

Proposition \ref{prop:corr} formally establishes that the perfect temporal correlation of the FF model is an idealized case existing only under a strict spectral condition. The proof guarantees that any stationary model with advection that deviates even slightly from the FF's rigid structure will inherently exhibit temporal decorrelation along a propagation path.
Among the models in this class, we build one where the Dirac delta is substituted with a decaying function that distributes power to a broader range of frequencies around the central frequency $\omega = -\bk^\top \bv$. This choice, inspired by seasonally persistent processes \citep{andvel1986long,gray1989generalized} and models of geophysical phenomena \citep{stein1999interpolation}, has the advantage of preserving the interpretability of the FF while being more physically plausible. The following definition introduces a unified model.

\begin{definition}[Damped FF]
\label{def:damped_ff_unified}
A {\em Damped Frozen Field} is a zero-mean Gaussian process characterized by a persistence parameter $\alpha > 1/2$ and a damping parameter $\beta \geq  0$, whose spectrum is defined as
\begin{equation}
\label{eq:damped_ff_spectrum_unified}
   S_{ZZ}(\mathbf{k},\omega) = S_{XX}(\mathbf{k}) \cdot B(\bk, \omega ; \bv, \beta)^{-\alpha},
\end{equation}
where 
\begin{align*}
     B(\bk, \omega ; \bv, \beta) = (\omega + \mathbf{k}^\top \mathbf{v})^2 + (\beta \omega)^2.
\end{align*}
\end{definition}

This model captures the interplay between advection and physical decorrelation through its two key parameters. The persistence parameter, $\alpha$, controls the strength of the process's memory along the advective path. The condition $\alpha > 1/2$ ensures that the spectral density has finite 
integral over frequency. This is because, as $|\omega| \to \infty$, we have
$B(\mathbf{k}, \omega; \mathbf{v}, \beta)\sim \omega^2(1 + \beta^2)$,
so that $S_{ZZ}(\mathbf{k},\omega) \sim |\omega|^{-2\alpha}$. The integral 
$\int_{\mathbb{R}} |\omega|^{-2\alpha} d\omega$ converges if and only if 
$2\alpha > 1$, yielding the requirement $\alpha > 1/2$. Under such parameter restriction, the model describes a system where energy decays at higher frequencies, which is a necessary condition for a well-behaved physical process. The damping parameter $\beta \geq 0$ prevents the singularity of a FF at zero frequency, and governs the geometric shape of the spectral decay around the advection line.
Furthermore, for the total variance $c_{ZZ}(0,0) = \iiint S_{ZZ}(\mathbf{k}, \omega) \, \mathrm{d}\omega \, \mathrm{d}^2\mathbf{k}$ to be finite, the spatial spectrum $S_{XX}(\mathbf{k})$ must also be integrable over wavenumber $\mathbf{k}$, which is a standard assumption for well-defined spatial random fields.

The following result shows that the spectrum of the Damped FF corresponds to a valid covariance function.
\begin{proposition}
\label{prop:cov_from_sptr_unified}
The spectrum $S_{ZZ}(\mathbf{k},\omega)$ specified in Definition \ref{def:damped_ff_unified} corresponds to a valid, positive definite covariance function.
\end{proposition}
\begin{proof}
This follows from Bochner's theorem \citep{priestley1981spectral}. The spectrum $S_{ZZ}(\mathbf{k},\omega)$ is non-negative, as it is the product of the non-negative spatial spectrum $S_{XX}(\mathbf{k})$ and the strictly positive damping factor $B(\bk, \omega ; \bv, \beta)^{-\alpha}$. Since the Fourier transform establishes a one-to-one correspondence between non-negative spectra and positive definite covariance functions, $c_{ZZ}(\mathbf{h},\tau)$ is guaranteed to be valid.
\end{proof}
This spectral construction admits a rigorous interpretation within the theory of SPDE-based stationary random fields proposed by \cite{carrizo2022}. In their framework, the spatio-temporal spectral density is derived from the symbol function $g(\bk, \omega)$ of a differential operator. Our damping term $B(\bk, \omega ; \bv, \beta)^{-\alpha}$ plays an analogous role to the inverse of the squared symbol function $|g(\bk, \omega)|^{-2}$ in their Theorem 1. Consequently, the Damped FF can be viewed as the stationary solution to a fractional evolution SPDE driven by a source term with spatial spectrum $S_{XX}$. This connects our spectral approach to the advection-diffusion SPDEs studied by \cite{carlotto2024}. However, while \cite{carlotto2024} rely on Finite Element discretizations to handle spatially varying parameters and scattered data, our spectral formulation leverages the Fast Fourier Transform for computationally efficient simulation on regular grids.

Finally, the simulation of the Damped FF is done directly with Algorithm \ref{alg:DH}, as this method can generate a realization from any valid, specified spectrum. It suffices to substitute the Damped FF spectrum from Equation \eqref{eq:damped_ff_spectrum_unified} into the algorithm, bearing in mind the practical need to define the spectrum on a discrete frequency grid. We include the algorithm in the supplementary materials.

\subsection{Role of the Parameters}
\label{sec:role_parms}

The spectrum in Equation \eqref{eq:damped_ff_spectrum_unified} is shaped by the term $B(\bk, \omega ; \bv, \beta)^{-\alpha}$. The power-law form of this term determines the smoothness and decay of temporal correlations, centered on the advection line $\omega = - \bk^\top \bv$. While this does not constitute a long-memory process in the strict sense of a spectral divergence \citep{robinson2020spatial}, this formulation allows for a tunable degree of {\em temporal persistence}.
In particular, the term $B(\bk, \omega ; \bv, \beta)$ acts as a filter that intrinsically links a wave's temporal evolution to its spatial wavenumber, which is a physically realistic assumption. The filter reshapes the initial spatial spectrum by amplifying the signal where $B(\bk, \omega ; \bv, \beta)<1$, and attenuating it where $B(\bk, \omega ; \bv, \beta)>1$. The geometry of this amplification region is primarily controlled by the damping parameter $\beta$, while the persistence parameter $\alpha$ dictates the intensity of the effect.

\begin{figure}[htb!]
\begin{center}
\includegraphics[width=\textwidth]{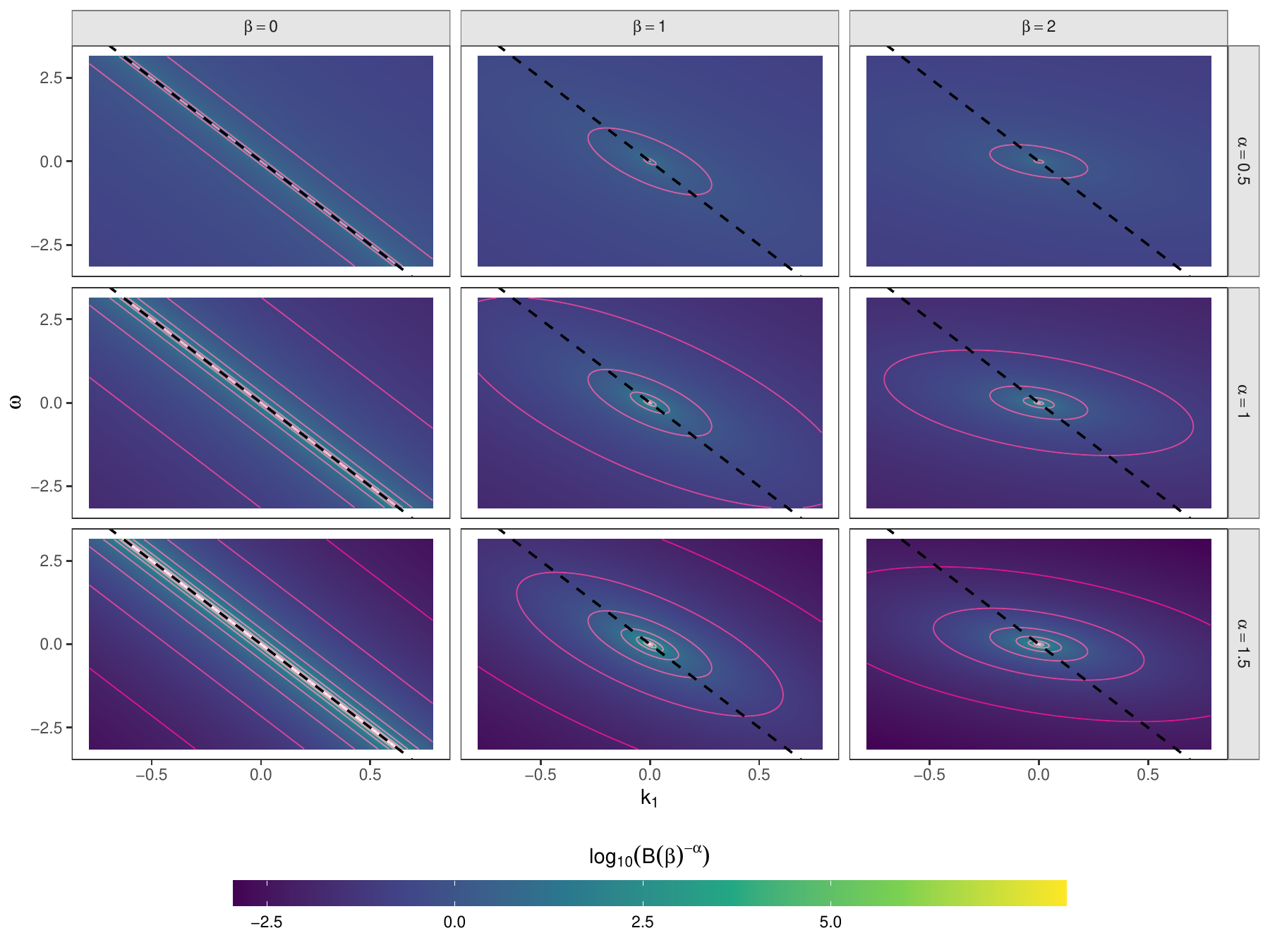}
\end{center}

\caption{Heatmap of the logarithm of $B(\bk, \omega ; \bv, \beta)^{-\alpha}$ for different choices of $\alpha$ (rows) and $\beta$ (columns). Level curves are shown in a gradient of pinks, where lightest correspond to highest values. The dashed black line correspond to $\omega = - \bk^\top \bv$. Note that $\alpha=0.5$ represents the theoretical limit of the model's validity, where the spectral density is at the threshold of integrability.}
\label{fig:alpha_beta_role}
\end{figure}

To illustrate this, we set $\bv=(5, 0)^\top$ and, by considering only the plane where $k_2=0$, we visualize the behavior of $B(\bk, \omega ; \bv, \beta)$ over the domain $k_1 \in [-\pi/4, \pi/4]$ and $\omega \in [-\pi, \pi]$. Figure \ref{fig:alpha_beta_role} illustrates the interplay between damping and persistence parameters, and the relative effects on $\log_{10}\left( B(\bk, \omega ; \bv, \beta)^{-\alpha}\right)$. As mentioned previously, $\beta$ governs the geometric shape of the persistent region in the frequency domain. When $\beta=0$, the filter creates an unbounded corridor of perfect persistence along the advection line $\omega = - \bk^\top \bv$. However, when $\beta>0$, the $(\beta\omega)^2$ term effectively dampens low temporal frequencies. This closes the corridor into a bounded, elliptical, region, meaning only a specific band of frequencies is allowed to persist over time.
In turn, the parameter $\alpha$ controls the degree of this persistence by dictating the filter's selectivity. A higher $\alpha$ makes the filter more selective, as it amplifies frequencies inside the persistent region defined by $\beta$, and attenuates those outside. This makes the process resemble a pure advective field. 
Thus, we conclude that the persistence parameter controls the long-range temporal memory of the process.

In summary, the damping and persistence parameters offer a flexible framework to model a rich variety of space-time processes by capturing the interplay between advection and decay of the correlation.

\section{Conclusions}
\label{sec:concl}

In this work, we have developed flexible and physically interpretable models for spatio-temporal phenomena governed by advection. Recognizing the limitations of the classical FF model, especially its assumption of perfect correlation along the advection path, we introduced two distinct generalization strategies: one based on enriching the velocity structure, and the other on relaxing the spectral shape.

Our first strategy led to the development of the Distributed FF and the Evolving FF models. The Distributed FF, by representing the process as a superposition of fields with different advective velocities, provides a natural framework for modeling phenomena driven by multiple transport influences. We demonstrated that this model is not only capable of producing complex dynamics, such as those of a tropical cyclone, but also that it provides a direct stochastic analogue to the advection-diffusion PDE. The Evolving FF, in contrast, allows the velocity to vary locally in space and time, offering a powerful tool for simulating processes within complex, non-uniform flow fields, like the evolving rainbands during Hurricane Florence. For both models, we derived key properties, including their propagation paths, and provided efficient simulation algorithms based on modifications of the FF simulation.

Our second strategy directly addressed the spectral properties of the advective process. We introduced the Damped FF model, which replaces the singular Dirac delta in the FF spectrum with a flexible, continuous filter. This filter is governed by a persistence parameter, $\alpha$, and a damping parameter, $\beta$, which together control the rate of decorrelation and the geometric shape of the spectral power around the advection line $\omega = -\mathbf{k}^\top\mathbf{v}$. We proved that this construction yields a valid covariance function and explored how the interplay between $\alpha$ and $\beta$ allows for the modeling of a rich class of processes that bridge the gap between pure advection and rapid decay.\\

The frameworks established in this paper open several promising directions for future investigation. While this work has focused on model definition and simulation, an important avenue for future research is the development of robust statistical methods for fitting these models to observational data. Each of the three proposed models naturally suggests specific estimation strategies tailored to its mathematical structure. For the Distributed FF, the explicit analytical form of the covariance function in Equation \eqref{eq:cov_distFF} enables classical spatial statistics approaches. Minimum contrast estimation \citep{Cressie2011} could be employed by matching the theoretical covariance to an empirical estimate, with parameters governing the velocity distribution optimized to minimize a distance criterion. Alternatively, composite likelihood methods \citep{varin2011} offer a computationally tractable approach for large datasets by constructing a pseudo-likelihood from pairwise or block contributions, circumventing the need to evaluate the full joint density. For the Evolving FF, the central inferential challenge is recovering the latent velocity field $\mathbf{v}(\mathbf{s},t)$ from observed realizations of the process. This problem has a direct analogue in computer vision, where optical flow methods (see for instance \cite{horn1981}) estimate motion fields by minimizing discrepancies between consecutive frames subject to smoothness constraints. This approach was used in the application to Hurricane Florence data. Within a parametric framework, another strategy is to represent the velocity field using a tensor product of known basis functions, specifically spatial and temporal basis functions. Then, the basis coefficients become the target of the estimation. This reduces the infinite-dimensional problem to a finite-dimensional one, amenable to penalized least squares or Bayesian inference with appropriate priors on the coefficients \citep{wood2017generalized, ramsay2005functional}. Moreover, physical constraints, such as incompressibility ($\nabla \cdot \mathbf{v} = 0$), can be incorporated through the choice of penalty terms \citep{corpetti2002}. Finally, the Damped FF is particularly amenable to frequency-domain inference due to its explicit spectral density in Equation \eqref{eq:damped_ff_spectrum_unified}. Whittle likelihood estimation \citep{whittle1954} would provide a natural framework, approximating the log-likelihood as a function of the periodogram and the theoretical spectrum. This approach is computationally attractive because both the periodogram and the parametric spectrum can be evaluated efficiently via the Fast Fourier Transform, making it feasible to estimate the persistence parameter $\alpha$, the damping parameter $\beta$, and the advection velocity $\mathbf{v}$ even for large spatio-temporal datasets. Recent advances in debiased Whittle estimation \citep{guillaumin2019debiased} address edge effects and improve finite-sample performance, which would be particularly relevant given the bounded domains typical in applications. Second, many environmental systems involve multiple, co-evolving variables, like temperature and pressure, and pollutants and wind velocity. Extending our models to a multivariate setting, where several fields are jointly advected and interact, is a critical next step. The multivariate FF model of \cite{salvana2022jasa} provides a foundational starting point for developing multivariate versions of the Distributed, Evolving, and Damped FF models.  Finally, the Damped FF model introduces a structure of tunable persistence centered on the advection path. A deeper theoretical investigation could formally connect this model to the theories of fractional dynamics and regularity found in the fractional Brownian motion literature (e.g., \cite{flandrin1989, qian1998two, stein2002fast, lilly2017fractional, didier2018domain}). For instance, it would be valuable to understand the interaction between the fractional characteristics of the initial spatial field $X_S(\bs)$ and the regularity properties induced by the advective spectral filter.

\section{Declaration of Interest}
None.

\section{Acknowledgments}
Sofia Olhede would like to thank the European Research Council under Grant CoG 2015-682172NETS, within the Seventh European Union Framework Program.

\bibliographystyle{apalike}
\bibliography{bibliography.bib}

\newpage
\appendix

\section{Algorithms}
This section outlines the simulation algorithms for the models presented in the main paper. Algorithm \ref{alg:sim_DistrFF} shows how to simulate a Distributed Frozen Field as a convex combination of $n_v$ simulated Frozen Fields; each component $i=1,\dots,n_v$ is advected with a unique velocity $\bv_i = \mathbf{M}_i\bv$ and assigned a weight $p_i = p(\mathbf{M}_i\bv)$. For the Evolving Frozen Field, Algorithm \ref{alg:sim_EvFF} adapts the simulation procedure of a Frozen Field by substituting the constant advection velocity with a vector field $\bv(\bs,t)$ that varies for each spatial location $\bs$ and time point $t$. Finally, Algorithm \ref{alg:sim_DampedFF_spectral} presents the spectral simulation of a Damped Frozen Field, which is generated from a spectrum constructed using the spatial field's spectrum $S_{XX}$, and a filter whose shape is determined by a persistence parameter $\alpha>1/2$ and a damping parameter $\beta \geq 0$.




\begin{algorithm}[H]
\caption{Simulation of a Distributed Frozen Field via Grid Shifting}\label{alg:sim_DistrFF}
\begin{algorithmic}[1]
\Require Spatial grid dimension $n$, number of time steps $T$, base velocity $\bv$, spatial covariance function $c_{XX}$.
\Require Number of components $n_v$, transformation matrices $\{\mathbf{M}_1, \dots, \mathbf{M}_{n_v}\}$, weights $\{p_1, \dots, p_{n_v}\}$.

\State Compute transformed velocities $\bv_i \gets \mathbf{M}_i \bv$ for $i=1, \dots, n_v$.
\State Compute max velocity component $v_{\text{max}} \gets \max_{i \in \{1,\dots,n_v\}, j \in \{1,2\}} \{ |(\bv_i)_j| \}$.
\State Compute extended grid dimension $N \gets n + T v_{\text{max}}$.
\State Define extended grid $\mathcal{G}^{*} \gets [0,N] \times [0,N]$. 
\State Generate Gaussian random field $X_{S}^{\text{ext}}$ on $\mathcal{G}^{*}$ using $c_{XX}$. 
\State Define simulation grid $\mathcal{G} \gets [0,n] \times [0,n]$.
\State Initialize space-time field $Z \in \mathbb{R}^{(n+1) \times (n+1) \times (T+1)}$.

\For{each spatial location $\bs$ in $\mathcal{G}$}
    \State $Z(\bs, 0) \gets X_{S}^{\text{ext}}(\bs)$.
    \For{$t = 1$ to $T$}
        \State $\text{value} \gets 0$. 
        \For{$i = 1$ to $n_v$}
            \State $\text{value} \gets \text{value} + p_i \cdot X_{S}^{\text{ext}}(\bs - \bv_i t)$. 
        \EndFor
        \State $Z(\bs, t) \gets \text{value}$.
    \EndFor
\EndFor
\State \Return $Z$.
\end{algorithmic}
\end{algorithm}

\begin{algorithm}[H]
\caption{Simulation of an Evolving Frozen Field via Grid Shifting}\label{alg:sim_EvFF}
\begin{algorithmic}[1]
\Require Spatial grid dimension $n$, number of time steps $T$, velocity field $\bv(\bs, t)$, spatial covariance function $c_{XX}$.

\State Define simulation grid $\mathcal{G} \gets [0,n] \times [0,n]$.
\State Compute max velocity component $v_{\text{max}} \gets \max_{\bs \in \mathcal{G}, t \in \{1,\dots,T\}, j \in \{1,2\}} \{ |v_j(\bs, t)| \}$.
\State Compute extended grid dimension $N \gets n + T v_{\text{max}}$.
\State Define extended grid $\mathcal{G}^{*} \gets [0,N] \times [0,N]$. 
\State Generate Gaussian random field $X_{S}^{\text{ext}}$ on $\mathcal{G}^{*}$ using $c_{XX}$. 
\State Initialize space-time field $Z \in \mathbb{R}^{(n+1) \times (n+1) \times (T+1)}$.

\For{each spatial location $\bs$ in $\mathcal{G}$}
    \State $Z(\bs, 0) \gets X_{S}^{\text{ext}}(\bs)$. 
    \For{$t = 1$ to $T$}
        \State $\bv_{st} \gets \bv(\bs, t)$. 
        \State $Z(\bs, t) \gets X_{S}^{\text{ext}}(\bs - \bv_{st} t)$. 
    \EndFor
\EndFor
\State \Return $Z$.
\end{algorithmic}
\end{algorithm}

\begin{algorithm}[H]
\caption{Simulation of a Damped Frozen Field via its Spectrum}\label{alg:sim_DampedFF_spectral}
\begin{algorithmic}[1]
\Require Spatial grid dimension $n$, number of time steps $T$, advection velocity $\bv$.
\Require Spatial spectral density $S_{XX}$, persistence parameter $\alpha > 1/2$, damping parameter $\beta \geq 0$.

\State Define padded dimensions $n_{\text{pad}} \gets 2n$ and $T_{\text{pad}} \gets 2T$. 
\State Define wavenumber vector $\mathbf{k}_{\text{vec}}$ for a grid of size $n_{\text{pad}}$.
\State Define frequency vector $\omega_{\text{vec}}$ for a grid of size $T_{\text{pad}}$.
\State Initialize complex grid in frequency domain $\widetilde{Z} \in \mathbb{C}^{n_{\text{pad}} \times n_{\text{pad}} \times T_{\text{pad}}}$.

\For{each index $(i, j, t)$ in the padded grid dimensions}
    \State Get corresponding wavevector $\bk \gets (\mathbf{k}_{\text{vec}}[i], \mathbf{k}_{\text{vec}}[j])$ and frequency $\omega \gets \omega_{\text{vec}}[t]$.
    \State $S_{XX\_val} \gets S_{XX}(\bk)$.
    \State $S_{modulator} \gets \left[(\omega + \bk^\top \bv)^2 + (\beta \omega)^2\right]^{-\alpha}$. 
    \State $S \gets S_{XX\_val} \cdot S_{modulator}$. 
    \State Generate complex white noise $\gamma \sim \mathcal{CN}(0,1)$. 
    \State $\widetilde{Z}[i,j,t] \gets \sqrt{S} \cdot \gamma$. 
\EndFor

\State Perform inverse Fourier transform $Z_{\text{pad}} \gets \text{IFFT3}(\widetilde{Z})$. 
\State Take the real part $Z_{\text{pad}} \gets \Re(Z_{\text{pad}})$.
\State Crop to original dimensions $Z \gets Z_{\text{pad}}[0:n-1, 0:n-1, 0:T-1]$.

\State \Return $Z$.
\end{algorithmic}
\end{algorithm}

\section{Derivation of transport-diffusion PDE}
In this section we expand on the link between Distributed Frozen Field and the solution of a transport-diffusion PDE.

We recall that a Distributed Frozen Field $Z$ is the convex combination of Frozen Fields $\{Z_i \}_{i=1}^{n_v}$, i.e.
\begin{equation}
    Z(\bs,t) = \sum_{i=1}^{n_v} p_i Z_i(\bs,t) = \sum_{i=1}^{n_v}  p_i X_S \left(\bs - \bv_it\right),
    \label{eq:dff_model}
\end{equation}
where we called $\bv_i = \mathbf{M}_i \bv$, namely the velocity of field $i$, with corresponding weight $p_i = p(\mathbf{M}_i \bv)$. By definition,  each Frozen Field $Z_{i}$, $i=1,\dots, n_v$, satisfies
\begin{equation}
    \frac{\partial Z_i(\bs,t)}{\partial t} + \bv_i \cdot \nabla Z_i(\bs,t) = 0,
    \label{eq:pure_advection}
\end{equation}
with the initial condition $Z_i(\bs,0) = X_S(\bs)$. 

Computing the time derivative of $Z(\bs,t)$ in \eqref{eq:dff_model}, we get
\begin{align*}
    \frac{\partial Z (\bs,t)}{\partial t} = \sum_{i=1}^{n_v} p_i \frac{\partial Z_i (\bs,t)}{\partial t} = -\sum_{i=1}^{n_v} p_i \left(  \bv_i \cdot \nabla Z_i (\bs,t) \right),
\end{align*}
where the last step was uptained by plugging in \eqref{eq:pure_advection}.
Now, substitute $\bv_i = \overline{\bv} + \Delta\bv_i$:
\begin{equation}
\label{eq:DFF_mean_deltas}
    \frac{\partial Z(\bs,t)}{\partial t} =  - \overline{\bv} \cdot \left( \sum_{i=1}^{n_v} p_i \nabla Z_i(\bs,t) \right) - \sum_{i=1}^{n_v} p_i (\Delta\bv_i \cdot \nabla Z_i(\bs,t)).
\end{equation}  
Since $ \sum_{i=1}^{n_v} p_i \nabla Z_i(\bs,t) = \nabla Z(\bs,t)$ thanks to the definition \eqref{eq:dff_model}, we can rearrange \eqref{eq:DFF_mean_deltas} into
\begin{equation}
\label{eq:DFF_pre_PDE}
    \frac{\partial Z(\bs,t)}{\partial t} + \overline{\bv} \cdot\nabla Z(\bs,t) = - \sum_{i=1}^{n_v} p_i (\Delta\bv_i \cdot \nabla Z_i(\bs,t)).
\end{equation}

For each component $i=1,\dots,n_i$, 
$$(\nabla Z_i) \cdot \Delta\bv_i = \nabla \cdot (Z_i \Delta\bv_i) - Z_i (\nabla \cdot \Delta\bv_i),$$
and $\nabla \cdot \Delta\bv_i = 0$ since $\Delta\bv_i$ does not vary with $(\bs, t)$. Thus, \eqref{eq:DFF_pre_PDE} becomes
\begin{align*}
    \frac{\partial Z(\bs,t)}{\partial t} + \overline{\bv} \cdot\nabla Z(\bs,t) = - \nabla \cdot \left( \sum_{i=1}^{n_v} p_i\, Z_i(\bs, t) \Delta \bv_i \right),
\end{align*}
as shown in the paper.

\end{document}